\documentclass[12pt]{amsart}
\usepackage{ifthen,verbatim}
\usepackage{floatflt,graphicx,graphics}
\usepackage{subfig}

\usepackage{tikz} 
\usepackage{mathrsfs}
\usepackage{amsmath,amssymb,amsthm}
\usepackage{mathtools}
\usepackage{floatflt} 

\numberwithin{equation}{section}
\nonstopmode
\numberwithin{equation}{section}
\theoremstyle{plain}

\theoremstyle{definition}

\theoremstyle{remark}

\DeclarePairedDelimiter\floor{\lfloor}{\rfloor}

{\qed\bigskip}

\newcounter{alphabet}

\newcounter{minutes}
\divide\time by 60
\newcounter{hours}
\multiply\time by 60
\addtocounter{minutes}{-\time}

\begin{document}
\bibliographystyle{amsplain}
\title
{
Different coefficients for studying dependence
}

\def\thefootnote{}
\footnotetext{
\texttt{\tiny File:~\jobname .tex,
          printed: \number\year-\number\month-\number\day,
          \thehours.\ifnum\theminutes<10{0}\fi\theminutes}
}
\makeatletter\def\thefootnote{\@arabic\c@footnote}\makeatother

\author[O. Rainio]{Oona Rainio}

\begin{abstract}
Through computer simulations, we research several different measures of dependence, including Pearson's and Spearman's correlation coefficients, the maximal correlation, the distance correlation, a function of the mutual information called the information coefficient of correlation, and the maximal information coefficient (MIC). We compare how well these coefficients fulfill the criteria of generality, power, and equitability. Furthermore, we consider how the exact type of dependence, the amount of noise and the number of observations affect their performance. According to our results, the maximal correlation is often the best choice of these measures of dependence because it can recognize both functional and non-functional types of dependence, fulfills a certain definition of equitability relatively well, and has very high statistical power when the noise grows if there are enough observations. While Pearson's correlation does not find symmetric non-monotonic dependence, it has the highest statistical power for recognizing linear and non-linear but monotonic dependence. The MIC is very sensitive to the noise and therefore has the weakest statistical power.
\end{abstract}
\maketitle

\section{Introduction}

In the study of statistics, one very often needs to somehow measure the dependence between two variables to understand their behavior. It is useful to know if there is some relationship, how strong it is and if we can use it, for instance, to predict the future observations. Consequently, it is important to have a suitable coefficient that works as a measure of dependence.   

Several different options have been introduced for this exact purpose over the history. Already in the 19th century, Pearson's correlation coefficient was first defined to identify linear dependence between variables. Later, its definition was extended to create Spearman's correlation coefficient in 1904 by C. Spearman \cite{s04}, the maximal correlation in 1941 by H.\ Gebelein \cite{g41}, and the distance correlation in 2007 by G.J.\ Sz\'ekely et al.\ \cite{s07} so that also non-linear and non-monotonic dependence could be detected. The birth of C.\ Shannon's information theory \cite{s48} in the 1940s enabled measuring non-functional dependence by using the mutual information, as formulated in 1957 by E.H.\ Linfoot \cite{l57}, and yet another quantity named the maximal information coefficient (MIC) was proposed in 2011 by D.N.\ Reshef et al.\ \cite{r11}. Furthermore, there exist local measures of dependence, such as the correlation curve \cite{b93} and the local Gaussian correlation \cite{t13}, and dependence between random variables can be also described with a type of multivariate cumulative distribution function called a copula \cite{s59}.

It is important to note that the coefficients of a single number or index cannot fully reveal the real nature of the underlying dependence \cite{b09} but, given their simple expression, different correlation coefficients, mutual information, and the MIC are very useful and therefore interesting topics of study. However, the number of these coefficients brings forth the question about which one of them should be used in a given situation. In \cite{r59}, A. R\'enyi introduced seven fundamental properties for a measure of dependence, including symmetry, values ranging the interval $[0,1]$, and the value 0 meaning independence, but most of these requirements are trivially fulfilled by the aforementioned coefficients or their slightly modified versions. In order to compare them and determine which measure of dependence works the best, we use here the three following criteria, out of which the first and the third one were introduced in \cite{r11} and the second one is notably studied in \cite{k14}. 

Firstly, we need to consider the \emph{generality} of the measures of dependence because it is important that a chosen coefficient can be applied into different situations. Does our quantity only detect linear, monotonic, or functional dependence, or can it also recognize more complicated relationships between the variables? It must be taken into account whether the coefficient is designed for continuous or discrete variables, and how many observations it needs to work properly.

The other significant requirement is the \emph{power} of the coefficient. How effective the measure is when used in a statistical test to decide whether there is some association between the variables or not? Namely, we can use any of our measures to test a null hypothesis of no dependence between two variables by first choosing a suitable threshold value from data of independent variables so that the probability of rejecting a true null hypothesis is fixed and then compute the probability of rejecting a false null hypothesis with the chosen threshold and data of two dependent variables. It is known that the amount of statistical noise in the relationship affects the power of the coefficients and, in particular, the MIC has been criticized for having too low power in case of noisy data \cite{s14}.

The third criterion is the \emph{equitability} of the measures of dependence. Does the coefficient give similar values for such relationships that are based on different functions but have the same level of noise? Especially, this property was first attributed for the MIC in \cite{r11} but, according to \cite{k14}, it does not work as well as implied earlier. 

While each of the coefficients considered here has been already studied separately \cite{a15, k14, x16} and there is a survey article by D. Tj\o{}stheim et al. \cite{t22} about copulas and local measures of dependence, there is relatively little research comparing different non-local measures based only one coefficient. Our aim in this article is to fill this gap by studying Pearson's and Spearman's correlation coefficients, the maximal correlation, the distance correlation, mutual information, and the MIC together. To find out if there is some coefficient that detects dependence always better than the others, we study them experimentally through several simulations implemented with the programming language R.

The structure of this article is as follows. First, we define of all the measures of dependence studied here and explain the methods for their computation in Section \ref{s2}. In Section \ref{s3}, we introduce our models and check what kind of values our coefficients give for them. Then, in Section \ref{s4}, we compare the power of our coefficients by also considering how it is affected by certain elements, such as the exact type of dependence, the amount of noise, and the number of observations. Finally, in Section \ref{s5}, we study the equitability of the coefficients under different functional relationships.

\section{Preliminaries}\label{s2}

Let us first define all the measures of dependence and show how they can be computed with the programming language R. If we have observations $(x_i,y_i)$, $i=1,...,n$, from two variables $X$ and $Y$, we can estimate the correlation between these variables by computing \emph{Pearson's correlation coefficient} \cite[(4), p.\ 3868]{x16}
\begin{align}
r=\frac{\sum^n_{i=1}(x_i-\overline{x})(y_i-\overline{y})}{\sqrt{\sum^n_{i=1}(x_i-\overline{x})^2\sum^n_{i=1}(y_i-\overline{y})^2}}\in[-1,1],    
\end{align}
where $\overline{x}$ and $\overline{y}$ denote the means of vectors $(x_1,...,x_n)$ and $(y_1,...,y_n)$, respectively. This coefficient was designed for measuring linear dependence between two variables whose marginal distributions are assumed to be normal, but it can also recognize non-linear dependence as long as it is monotonic.  

One of the most well-known alternatives for Pearson's correlation coefficient is \emph{Spearman's correlation coefficient} $r_s$, which is found for $n$ paired observations $(x_i,y_i)$ by first converting them into their rank numbers and then calculating Pearson's correlation coefficient of these ranks \cite[p.\ 3869]{x16}. Spearman's coefficient is also from the interval $[-1,1]$ but, compared to Pearson's coefficient, it suits better for such situations where the dependence is non-linear but monotonic or the variables are not normally distributed. Still, neither Pearson's nor Spearman's correlation coefficient is a good choice when the relationship between the variables is non-monotonic. 

However, we can use the \emph{maximal correlation} \cite[(1), p.\ 27]{a15}
\begin{align}\label{q_mcc}
\rho_{\max}=\sup\{\rho(f_0(X);f_1(Y))\}\in[0,1]    
\end{align}
to measure all types of functional dependence, regardless of if they are monotonic or not. Above, the supremum is taken over all the real-valued functions $f_0,f_1$ defined for the values of the variables $X$ and $Y$, respectively, such that $\text{E}(f_0(X))=\text{E}(f_1(Y))=0$ and $\text{E}(f_0(X)^2)=\text{E}(f_1(Y)^2)=1$. The notation $\rho(;)$ means here the population correlation, which is can be estimated from the data by computing Pearson's coefficient $r$.

Another measure of dependence based on the definition of correlation is the \emph{(sample) distance correlation}
\begin{align}\label{q_cdist}
\rho_{\text{dist}}=
\sqrt{\frac{\mathcal{V}^2_n(X;Y)}{\sqrt{\mathcal{V}^2_n(X)\mathcal{V}^2_n(Y)}}}\in[0,1],
\end{align}
where, for $n$ paired observations $(x_i,y_i)$ from the variables $X$ and $Y$,
\begin{align*}
&\mathcal{V}^2_n(X;Y)
=\frac{1}{n^2}\sum^n_{j=1}\sum^n_{k=1}A_{j,k}B_{j,k},\quad
\mathcal{V}^2_n(X)=\mathcal{V}^2_n(X;X),\quad
\mathcal{V}^2_n(Y)=\mathcal{V}^2_n(Y;Y),\\
&A_{j,k}=|x_j-x_k|-\frac{1}{n}\sum^n_{l=1}|x_j-x_l|-\frac{1}{n}\sum^n_{l=1}|x_k-x_l|+\frac{1}{n^2}\sum^n_{l=1}\sum^n_{h=1}|x_l-x_h|,\quad\text{and}\\
&B_{j,k}=|y_j-y_k|-\frac{1}{n}\sum^n_{l=1}|y_j-y_l|-\frac{1}{n}\sum^n_{l=1}|y_k-y_l|+\frac{1}{n^2}\sum^n_{l=1}\sum^n_{h=1}|y_l-y_h|.
\end{align*}
This coefficient is much newer than the previous ones and should be able to recognize different functional relationships. Note that if the denominator in \eqref{q_cdist} is 0, we simply set $\rho_{\text{dist}}=0$.

A slightly different way to identify dependence is compute the \emph{mutual information} between variables $X$ and $Y$, which is defined as a sum
\cite[p.\ 431]{v09}
\begin{align}
I(X;Y)=\sum_i\sum_j p(x_i,y_j)\log_2\left(\frac{p(x_i,y_j)}{p(x_i)p(y_j)}\right)\in[0,\infty) 
\end{align}
for discrete random variables $X$ and $Y$ with values $x_i$ and $y_j$, and as an integral \cite[(14), p. 88]{l57}
\begin{align}
I(X;Y)=\int_{x\in\mathcal{X}}\int_{y\in\mathcal{Y}} p(x,y)\log_2\left(\frac{p(x,y)}{p(x)p(y)}\right)dxdy\in[0,\infty).    
\end{align}
for continuous random variables $X$ and $Y$ with value sets $\mathcal{X}$ and $\mathcal{Y}$. While the exact value of the mutual information is often quite difficult to find because it requires knowing the probability distribution function $p$, this quantity can be estimated by dividing the domain into small bins and then using the so-called \emph{naive estimate} \cite[(6), p.\ 3356]{k14}
\begin{align}
I_\text{naive}(X;Y)=\sum_{\widetilde{x},\,\widetilde{y}}\hat{p}(\widetilde{x},\widetilde{y})\log_2\left(\frac{\hat{p}(\widetilde{x},\widetilde{y})}{\hat{p}(\widetilde{x})\hat{p}(\widetilde{y})}\right),   
\end{align}
where $\hat{p}(\widetilde{x},\widetilde{y})$ is the fraction of data points inside one bin. The mutual information tells us the expected amount of information that the observations of one variable give about the other variable, and this measure therefore describes also non-functional relationships.   

By denoting the estimate of the mutual information found with the bins of a rectangular $n_x\times n_y$-grid $G$ by $I_G(X;Y)$, we can write the definition of the \emph{maximal information coefficient (MIC)} as \cite[(7), p.\ 3356]{k14}
\begin{align}
\text{MIC}(X;Y)=\max_{n_x\times n_y}\frac{\max_G I_G(X;Y)}{\log(\min\{n_x,n_y\})}\in[0,1].    
\end{align}
Here, the value of the product $n_x\times n_y$ has usually some upper bound, such as $B(n)=n^{0.6}$, where $n$ is the number of paired observations. Clearly, the MIC is a non-parametric measure of dependence between the variables $X$ and $Y$ and, since its definition is based on that of the mutual information, it should also be able to detect both functional and non-functional dependence.

One of the issues when comparing these measures of dependence is that they are defined on different intervals. Here, we are interested in such a coefficient whose value is 0 if the variables $X$ and $Y$ are independent, 1 if one of these variables fully determines the values of the other, and some number from the interval $(0,1)$ if there is a relationship between $X$ and $Y$ so that this value decreases as the amount of noise in the data increases. The maximal correlation, the distance correlation and the MIC already fulfill this condition, but we will consider below only the absolute values of both Pearson's and Spearman's correlation coefficient to deal with their values indicating negative correlation. Furthermore, because the mutual information is measured in bits and has sometimes values greater than 1, we consider the \emph{information coefficient of correlation} \cite[(13), p. 88]{l57}
\begin{align}\label{q_r1}
r_1=\sqrt{1-e^{-2\cdot I(X;Y)}}\in[0,1],    
\end{align}
which was introduced in 1957 by H.E. Linfoot so that the value of the mutual information could be interpreted better.

Let us yet briefly introduce the methods of computation used in our simulations. Firstly, Pearson's correlation coefficient can be computed with the base R-function \texttt{cor} and this same function also returns Spearman's coefficient if we choose value "spearman" for its parameter "method". The maximal correlation is found by first maximizing the linear correlation with the alternative conditional expectations algorithm \texttt{ace} from the package \texttt{acepack} and then using the function \texttt{cor}. The distance correlation can be computed with the function \texttt{dcor} from the package \texttt{energy}. The coefficient $r_1$ is obtained by first discretizing the data with \texttt{discretize} from the package \texttt{infotheo}, estimating the mutual information the function \texttt{mutinformation} from the same package and just applying the formula in \eqref{q_r1} in R. Finally, the MIC is computed with the function \texttt{mine} from the package \texttt{minerva}. We use here default settings for each function and more details can be found in the manuals of these R-packages.




\section{Generality for different types of dependence}\label{s3}

\begin{figure}[!tbp]
  \centering
  \subfloat[Linear]{\includegraphics[width=0.231\textwidth]{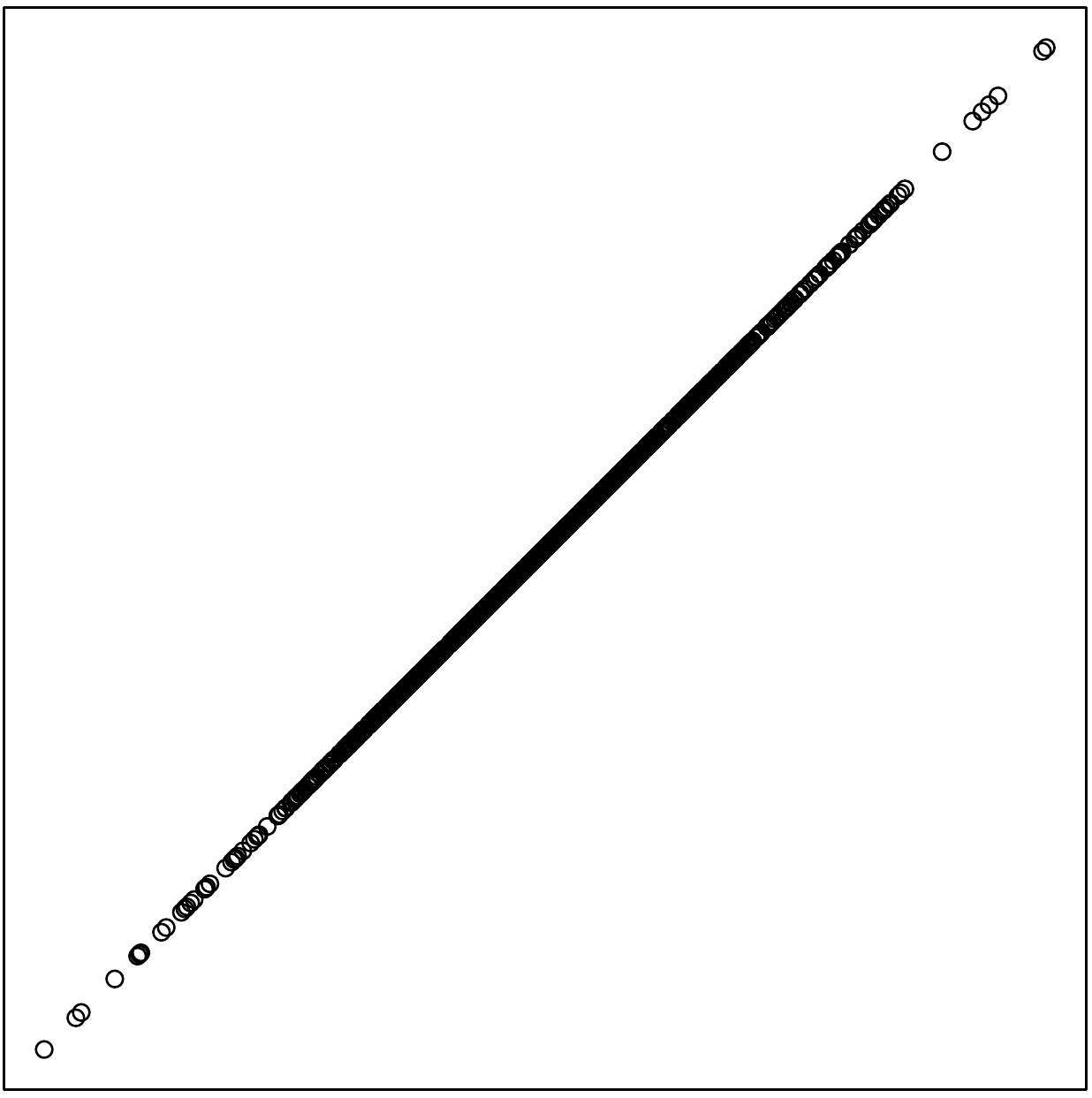}\label{fig311}}
  \hfil
  \subfloat[Logarithm]{\includegraphics[width=0.231\textwidth]{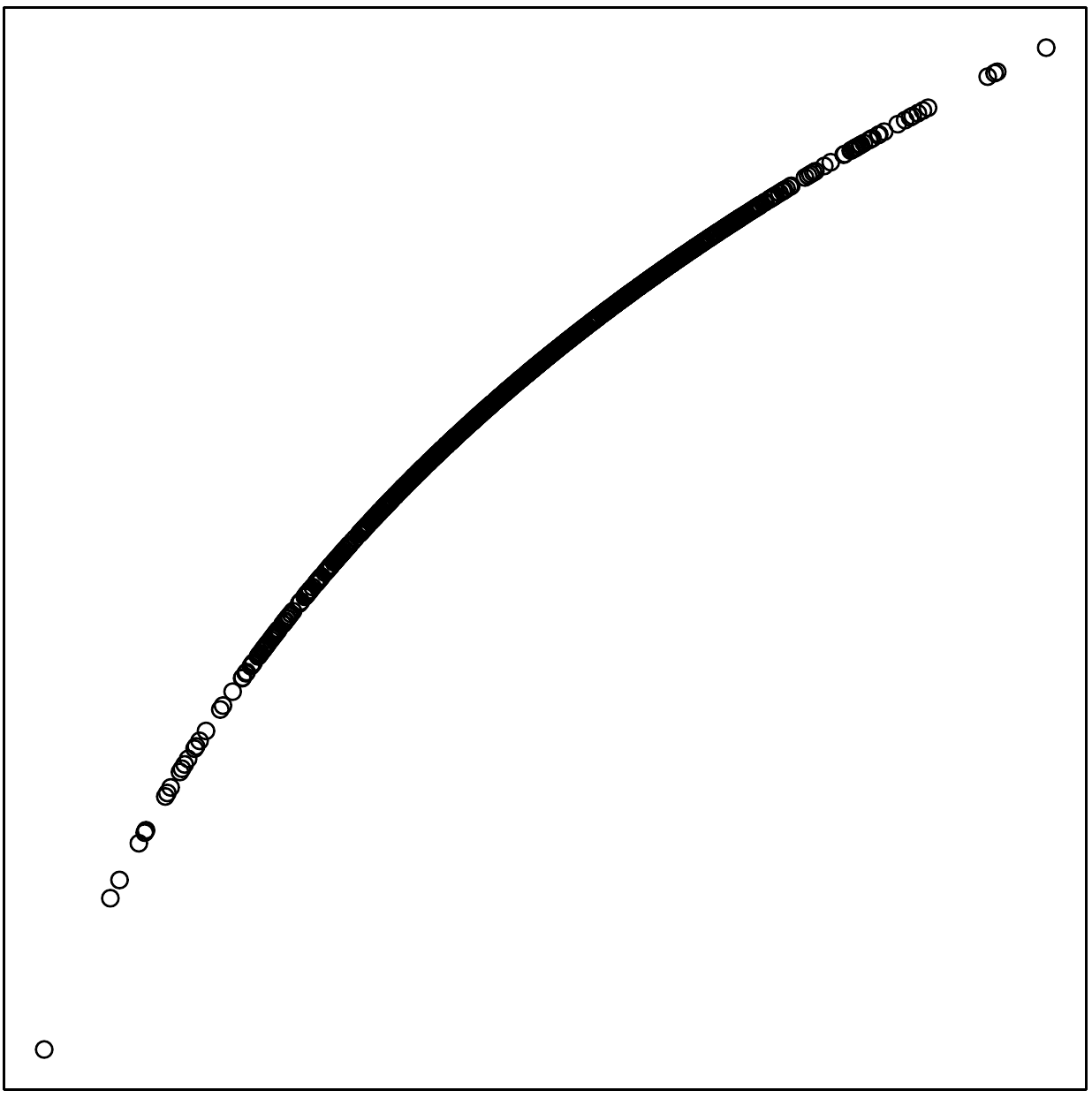}\label{fig312}}
  \hfil
  \subfloat[Cubic]{\includegraphics[width=0.231\textwidth]{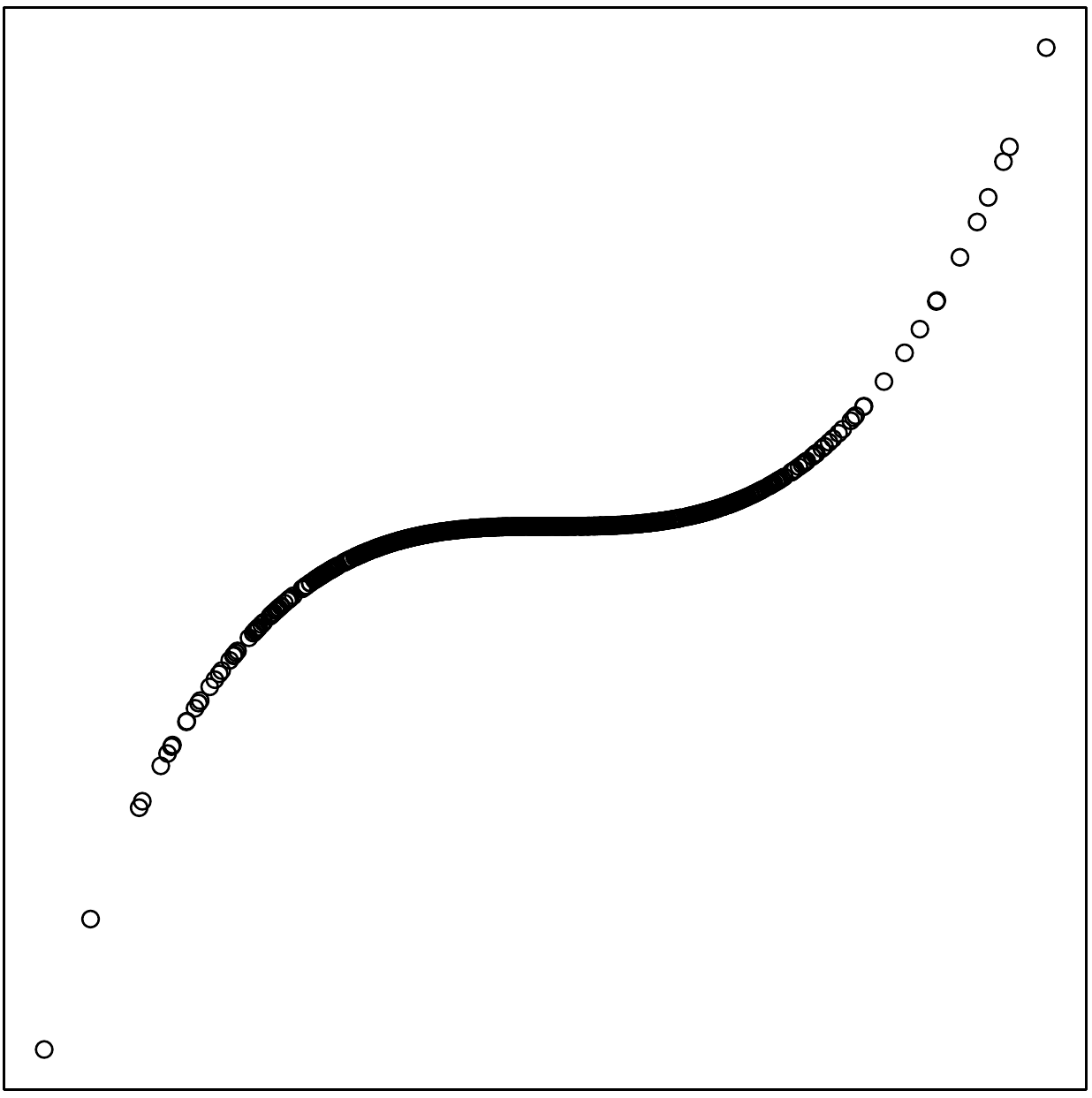}\label{fig313}}
  \\
  \subfloat[Quadratic]{\includegraphics[width=0.231\textwidth]{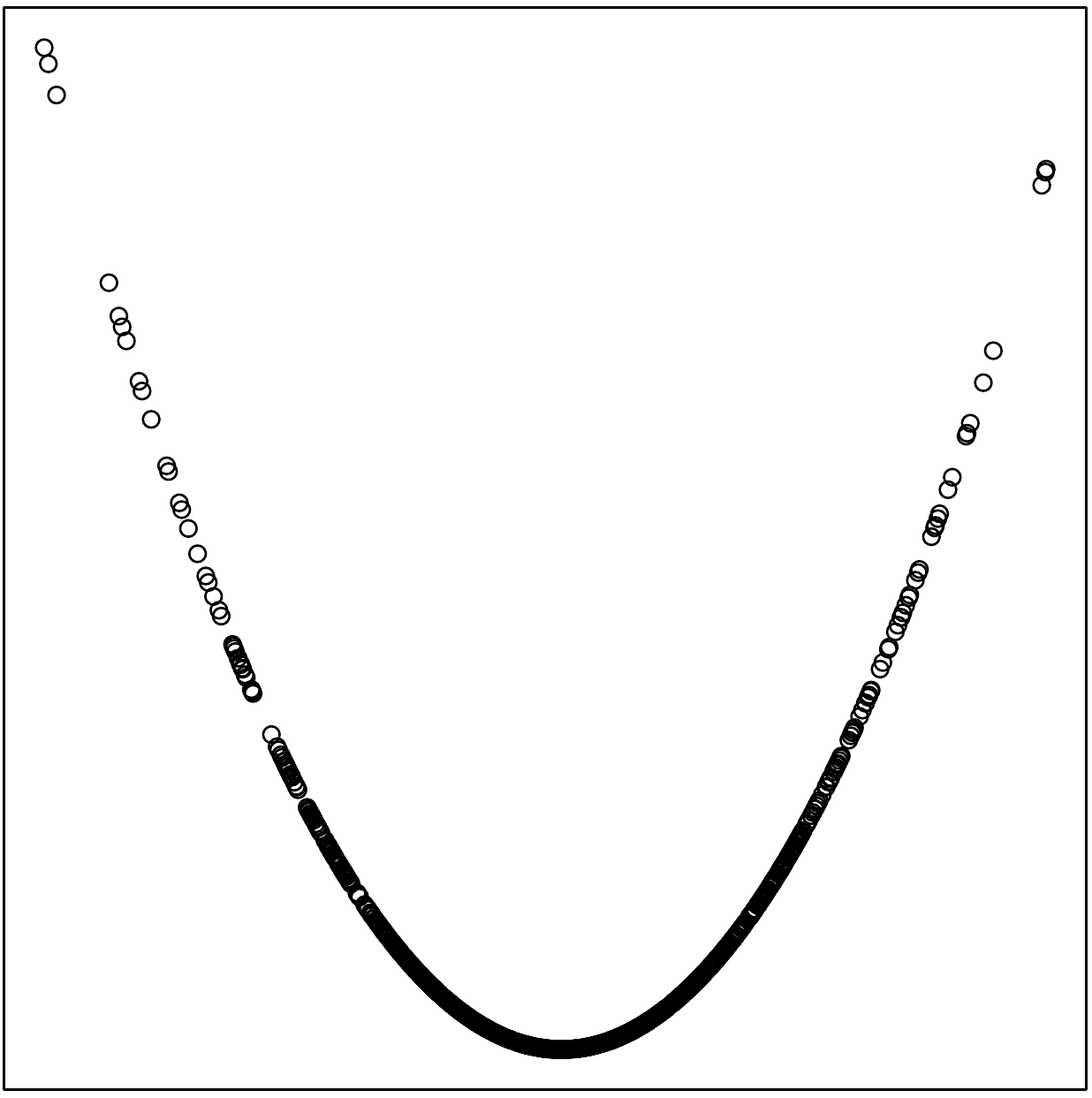}\label{fig314}}
  \hfil
  \subfloat[Sinusoidal]{\includegraphics[width=0.231\textwidth]{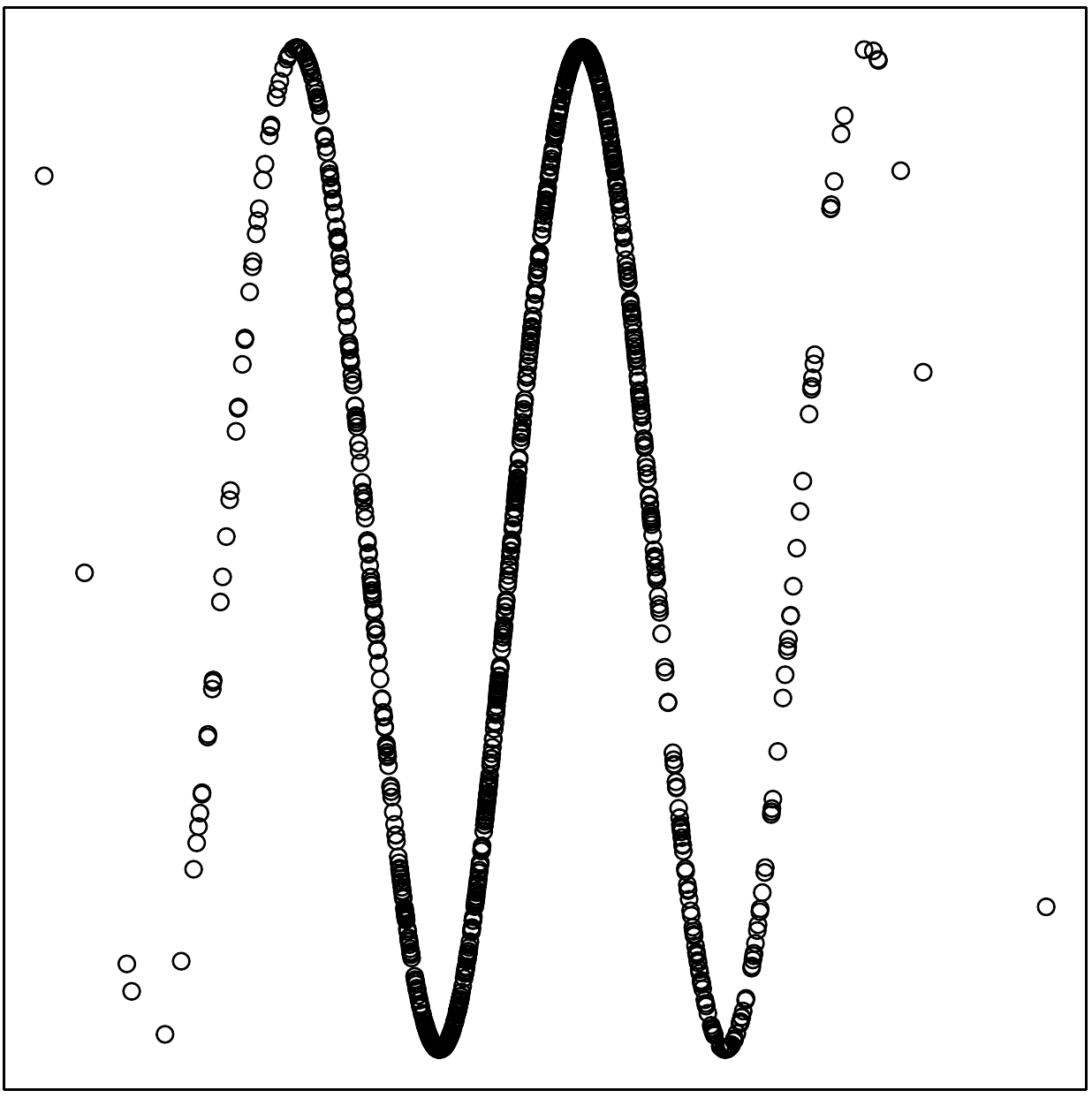}\label{fig315}}
  \hfil
  \subfloat[Piecewise]{\includegraphics[width=0.231\textwidth]{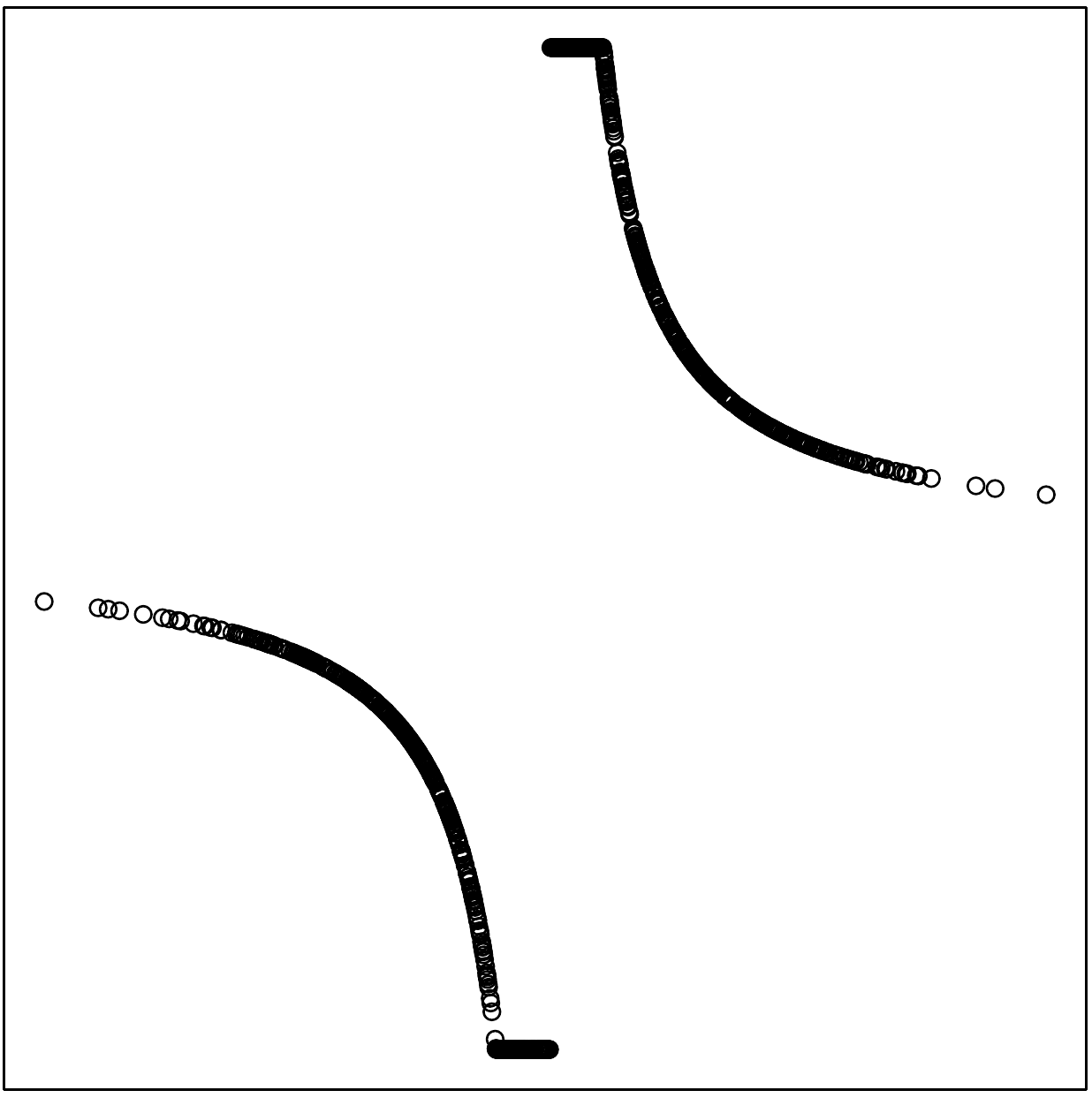}\label{fig316}}
  \\
  \subfloat[Cross]{\includegraphics[width=0.231\textwidth]{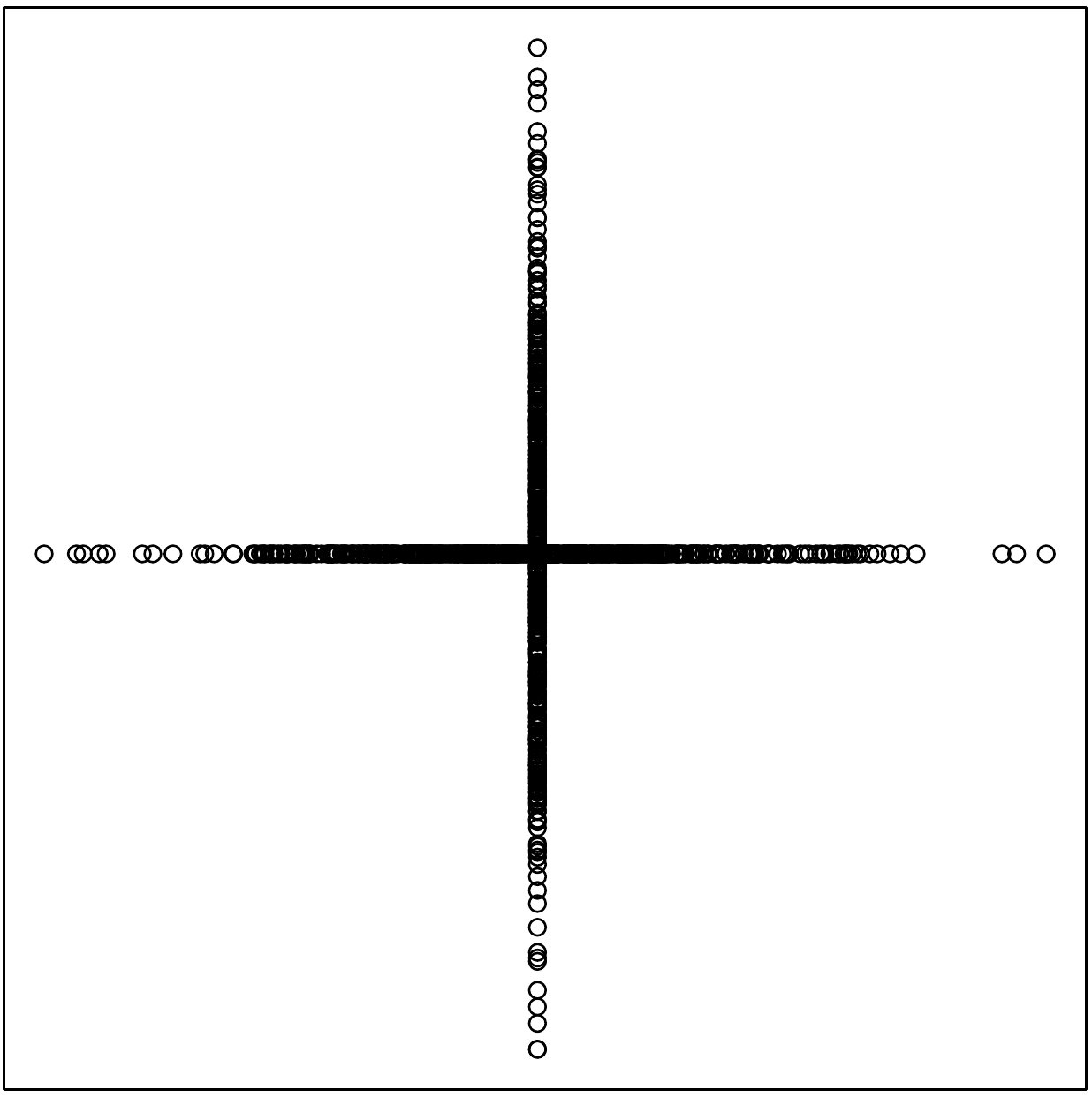}\label{fig317}}
  \hfil
  \subfloat[Circular]{\includegraphics[width=0.231\textwidth]{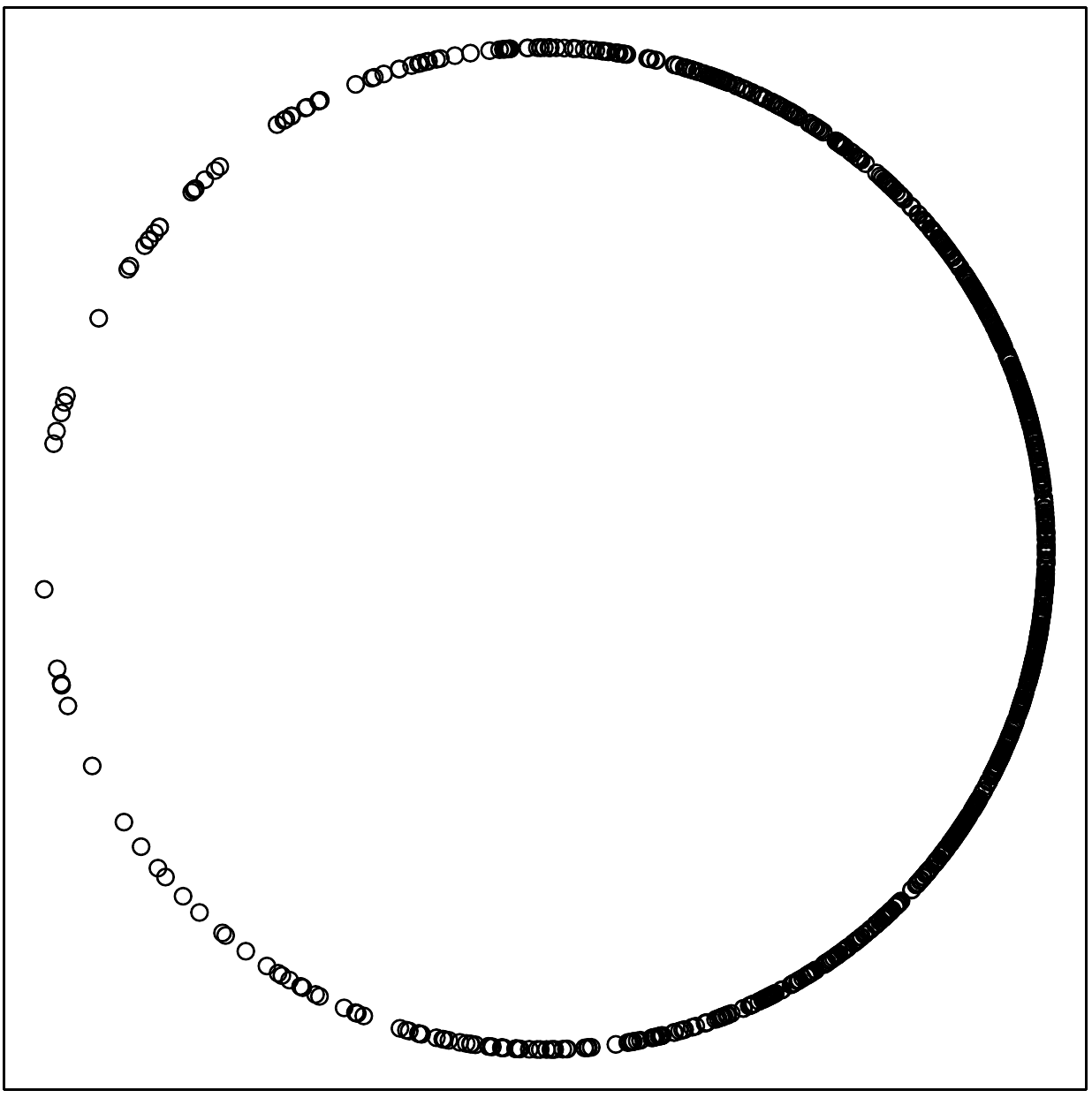}\label{fig318}}
  \hfil
  \subfloat[Checkers]{\includegraphics[width=0.231\textwidth]{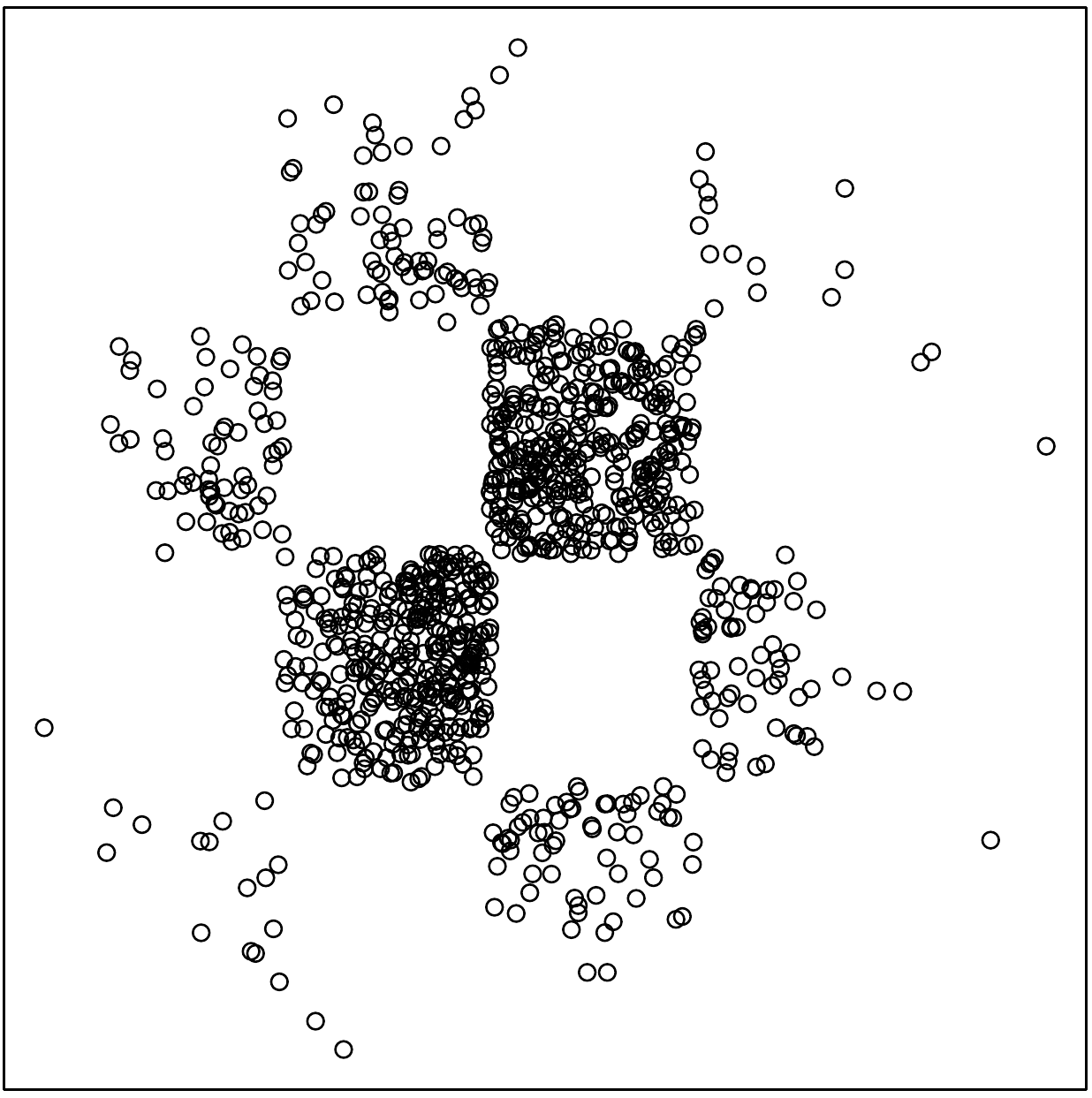}\label{fig319}}
  \caption{Scatter plots of one simulation from the models \eqref{q_funcmodel}-\eqref{q_chssmdl} with $\sigma=0$ and $n=1000$.}
  \label{fig3}
\end{figure}

In this section, we define nine different models of dependence, which can be seen from Figure \ref{fig3}. For each type of dependence, we study the values of six different measures introduced in the previous section. The models below are built by generating observations from the normal distribution for the explanatory variable, but they can be easily redefined for some other marginal distribution. 

In our simulations of functional dependence, the observations $i=1,...,n$ of the variables $X$ and $Y$ are generated according the model
\begin{align}\label{q_funcmodel}
x_i\sim N(0,1),\quad
y_i=f_j(x_i)+\epsilon_i,\quad
\epsilon_i\sim N(0,\sigma^2),
\end{align}
in which the function $f_j$ is either the linear, logarithmic, cubic, quadratic, sinusoidal, or piecewise function, defined as
\begin{align*}
&f_1(x)=x,\quad
f_2(x)=5\ln(|x+5|),\quad
f_3(x)=0.3x^3,\quad
f_4(x)=0.7x^2,\quad
f_5(x)=1.3\sin(3x),\\
&f_6(x)=\min\{\max\{1\slash x,-3\},3\}.
\end{align*}
We also compute our coefficients for three non-functional models of dependence, including the cross-shaped dependence
\begin{align}\label{q_crossmdl}
\begin{split}
&x_i\sim N(0,1),\quad
y_i\sim N(0,(\sigma\slash3)^2)
\quad\text{for }i=1,...,\floor{n\slash2},\\
&x_i\sim N(0,(\sigma\slash3)^2),\quad
y_i\sim N(0,1)
\quad\text{for }i=\floor{n\slash2}+1,...,n,
\end{split}
\end{align}
the circular dependence
\begin{align}
(x_i,y_i)\in\{(h_i\cos(k_i),h_i\sin(k_i))\text{ }|\text{ }h_i\sim N(1,(\sigma\slash7)^2),k_i\sim N(0,1)\},\quad
i=1,...,n,
\end{align}
and the checkerboard dependence
\begin{align}\label{q_chssmdl}
\begin{split}
&x_i=k_{i0},\quad y_i=k_{i1}+\epsilon_i,\quad\epsilon_i\sim N(0,(\sigma\slash2)^2),\quad i=1,...,n,\quad\text{where}\\
&
\begin{pmatrix}
k_{i0}\\
k_{i1}
\end{pmatrix}
\in
\left.\left\{
\begin{pmatrix}
k_0\\
k_1
\end{pmatrix}
\sim
N\left(
\begin{pmatrix}
0\\
0
\end{pmatrix}
,
\begin{pmatrix}
1 & 0\\
0 & 1
\end{pmatrix}
\right)
\text{ }\right|\text{ }
\floor{0.7k_0}-\floor{0.7k_1}\equiv 0\pmod{2}
\right\}.
\end{split}
\end{align}
These models have been created so that the amount of statistical noise in the data can be added by increasing the value of the parameter $\sigma>0$ in all the models except the cross-shaped model \eqref{q_crossmdl}, where the amount of noise is increasing with respect to $\sigma\in[0,3]$, decreasing with respect to $\sigma\geq3$, and the data comes from two independent, normally distributed variables if $\sigma=3$.

First, let us consider the noiseless versions of these models with 1000 observations to see how our coefficients recognize different types of dependence without any disrupting factors. For each model, we compute the average values of the coefficients $|r|$, $|r_s|$ $\rho_{\text{max}}$, $\rho_{\text{dist}}$, $r_1$, and MIC in 1000 simulations with $n=1000$ and $\sigma=0$. The results of this experiment are collected in Table \ref{t1}.

\begin{table}[ht]
    \centering
    \begin{tabular}{|l|l|l|l|l|l|l|}
       \hline
       Model & $|r|$ & $|r_s|$ & $\rho_{\text{max}}$ & $\rho_{\text{dist}}$ & $r_1$ & MIC \\
       \hline
       Linear & 1.000 & 1.000 & 1.000 & 1.000 & 0.994 & 1.000\\
       Logarithmic & 0.987 & 1.000 & 1.000 & 0.998 & 0.994 & 1.000\\
       Cubic & 0.779 & 1.000 & 0.995 & 0.854 & 0.994 & 1.000\\
       Quadratic & 0.056 & 0.033 & 1.000 & 0.542 & 0.969 & 1.000\\
       Sinusoidal & 0.049 & 0.123 & 0.984 & 0.359 & 0.919 & 1.000\\
       Piecewise & 0.441 & 0.504 & 0.979 & 0.735 & 0.973 & 1.000\\
       Cross-shaped & 0.001 & 0.001 & 0.931 & 0.328 & 0.800 & 0.631\\
       Circular & 0.027 & 0.032 & 0.995 & 0.411 & 0.958 & 0.996\\
       Checkerboard & 0.062 & 0.151 & 0.928 & 0.255 & 0.713 & 0.497\\
       \hline
    \end{tabular}
    \caption{The average values of the coefficients $|r|$, $|r_s|$, $\rho_{\text{max}}$, $\rho_{\text{dist}}$, $r_1$, and MIC in 1000 simulations of the models \eqref{q_funcmodel}-\eqref{q_chssmdl} with $n=1000$ and $\sigma=0$.}
    \label{t1}
\end{table}

From Table \ref{t1}, we see that Pearson's correlation coefficient $|r|$ has a value of 1 only for the linear dependence, Spearman's coefficient $|r_s|$ is 1 for all the monotonic relationships whereas the MIC is 1 for all functional models. Clearly, the two first coefficients cannot detect non-monotonic dependence properly and their values are very small for the symmetric models, like the cross-shaped, circular and quadratic types of dependence. Interestingly, the maximal correlation $\rho_{\text{max}}$ always has larger values than the coefficients $\rho_{\text{dist}}$ and $r_1$ and it also exceeds the MIC for the models \eqref{q_crossmdl} and \eqref{q_chssmdl}, even though the maximal correlation was designed only for identifying functional relationships.

\begin{table}[ht]
    \centering
    \begin{tabular}{|l|l|l|l|l|l|l|}
       \hline
       $\sigma$ & $|r|$ & $|r_s|$ & $\rho_{\text{max}}$ & $\rho_{\text{dist}}$ & $r_1$ & MIC \\
       \hline
       0.1 & 0.995 & 0.994 & 0.995 & 0.992 & 0.979 & 0.980\\
       0.5 & 0.895 & 0.885 & 0.895 & 0.862 & 0.873 & 0.663\\
       1 & 0.706 & 0.670 & 0.709 & 0.658 & 0.703 & 0.409\\
       3 & 0.316 & 0.303 & 0.321 & 0.287 & 0.384 & 0.181\\
       \hline
    \end{tabular}
    \caption{The average values of the coefficients $|r|$, $|r_s|$, $\rho_{\text{max}}$, $\rho_{\text{dist}}$, $r_1$, and MIC in 1000 simulations with $n=1000$ observations from the model \eqref{q_funcmodel} with the linear function $f_1(x)=x$, when the value of $\sigma$ varies.}
    \label{t2}
\end{table}

\begin{table}[ht]
    \centering
    \begin{tabular}{|l|l|l|l|l|l|l|}
       \hline
       $n$ & $|r|$ & $|r_s|$ & $\rho_{\text{max}}$ & $\rho_{\text{dist}}$ & $r_1$ & MIC \\
       \hline
       10 & 0.692 & 0.646 & 0.803 & 0.738 & 0.475 & 0.546\\
       100 & 0.706 & 0.684 & 0.739 & 0.666 & 0.659 & 0.508\\
       1000 & 0.706 & 0.670 & 0.709 & 0.658 & 0.703 & 0.409\\
       3000 & 0.707 & 0.690 & 0.707 & 0.657 & 0.711 & 0.370\\
       \hline
    \end{tabular}
    \caption{The average values of the coefficients $|r|$, $|r_s|$, $\rho_{\text{max}}$, $\rho_{\text{dist}}$, $r_1$, and MIC in 1000 simulations of the model \eqref{q_funcmodel} with the linear function $f_1(x)=x$ and $\sigma=1$, when the number $n$ of observations varies.}
    \label{t3}
\end{table}

By changing the values of the parameters $\sigma$ and $n$ in the simulations, we can see what kind of an impact the amount of noise and the number of observations, respectively, have on our measures of dependence. As we can see from Table \ref{t2}, the values of the MIC decrease notably faster than those of the other coefficients, when the noise levels grow. According to Table \ref{t3}, the correlation coefficients seem to decrease while the values of $r_1$ and the MIC increase with respect to $n$. Note here that, even though Pearson's coefficient can be defined for even just 3 observations, our methods of computation return 0 for the value of $r_1$ if $n\leq7$ and, similarly, the distance correlation cannot be computed either if $n\leq4$.

It can also be studied how our coefficients behave if we modify the model \eqref{q_funcmodel} so that the observations of the variable $X$ are generated from some distribution other than the standard normal distribution, such as the uniform, exponential or Poisson distribution. For instance, all the quantities give values close to 1 in case of the linear dependence, regardless of the exact marginal distribution of $X$, but the value of the distance correlation is greater for the sinusoidal model if we choose $X\sim\text{Pois}(3)$ instead. It must be noted that these changes obviously also affect the shape of the data, though, and such as noise parameter should be chosen that the amount of noise is proportional to the range of the variable $X$.

However, the values of our measures of dependence do not tell us very much without any additional information. In order to draw any conclusions whether a dependence in the data can be properly identified if, for instance, the MIC has a value of 0.3, we need to compare this result to the value of the coefficient computed from the data without any dependence. Consequently, we need to study here the power of our coefficients.

\section{Power for identifying dependence}\label{s4}

In this section, we study the power of six coefficients, including the absolute values of Pearson's and Spearman's correlation coefficients $r$ and $r_s$, the maximal correlation $\rho_{\text{max}}$, the distance correlation $\rho_{\text{dist}}$, the coefficient $r_1$, and the MIC. We apply the models \eqref{q_funcmodel}-\eqref{q_chssmdl} to create different types of dependence in our simulations. Furthermore, we consider how the amount of noise and the number of observations affect our results.

Recall that the power in a statistical test is the probability of rejecting a false null hypothesis. When studying the dependence between two variables, our null hypothesis is that there is no association between them, and we must therefore find out how likely it is to recognize the cases with some underlying dependence present. In order to measure this probability, we need to first decide the critical values of the coefficients which are used to decide if the null hypothesis is rejected or not with the significance level of $\alpha$. In other words, the power is of some coefficient $q$ is defined formally as the probability
\begin{align}
P(q(X,Y)>q_{\text{crit}}\text{ }|\text{ }X\not\perp Y)\quad\text{for}\quad\{q_{\text{crit}}\in[0,1]\text{ }|\text{ }P(q(X,Z)> q_{\text{crit}}\text{ }|\text{ }X\perp Z)=\alpha\}.
\end{align}

Consequently, let us compute the values of the coefficients $|r|$, $|r_s|$, $\rho_{\text{max}}$, $\rho_{\text{dist}}$, $r_1$, and MIC in 3000 simulations, each of which consists of $n=1000$ observations from two independent, similarly distributed variables. We have then some approximations of the distributions of these coefficients when the null hypothesis holds and, by taking the $(1-\alpha)$-quantiles from their histograms, we have estimates for their critical values for $\alpha$. Table \ref{t4} contains these estimates in the cases where both the variables follow the standard normal distribution $N(0,1)$ and $\alpha=1,5,10\%$.

\begin{table}[ht!]
    \centering
    \begin{tabular}{|l|l|l|l|l|l|l|}
       \hline
       $\alpha$ & $|r|$ & $|r_s|$ & $\rho_{\text{max}}$ & $\rho_{\text{dist}}$ & $r_1$ & MIC \\
       \hline
       1\% & 0.0827 & 0.0849 & 0.134 & 0.0934 & 0.301 & 0.152\\
       5\% & 0.0628 & 0.0619 & 0.116 & 0.0774 & 0.287 & 0.147\\
       10\% & 0.0517 & 0.0532 & 0.106 & 0.0706 & 0.278 & 0.143\\
       \hline
    \end{tabular}
    \caption{The critical values of the coefficients $|r|$, $|r_s|$, $\rho_{\text{max}}$, $\rho_{\text{dist}}$, $r_1$, and MIC estimated from 3000 simulations with $n=1000$ observations from two independent, normally distributed variables, when the significance varies.}
    \label{t4}
\end{table}

Now, we can estimate the power of our coefficients by computing what proportion of their values in 3000 simulations are above their critical values in Table \ref{t4}. In one experiment for all the models \eqref{q_funcmodel}-\eqref{q_chssmdl} with parameter choices $n=1000$, $\sigma=0.1$, and $\alpha=5\%$, it was observed that the powers of the coefficients $\rho_{\text{max}}$, $\rho_{\text{dist}}$, $r_1$, and MIC were 1 for all these models. The estimated powers of the absolute values of Pearson's and Spearman's correlation coefficients were 1 for all the monotonic relationships (the model \eqref{q_funcmodel} with $j=1,2,3$), but notably less than this for the other models. Especially, the powers of these two coefficients are close to 0 in case of symmetric non-monotonic dependence, like the cross-shaped dependence of model \eqref{q_crossmdl}.

\begin{figure}[!ht]
    \centering
    \begin{tikzpicture}
    \draw (0, 0) node[inner sep=0]
    {\includegraphics[trim={0cm 0cm 1cm 2cm},clip,width=11.5cm]{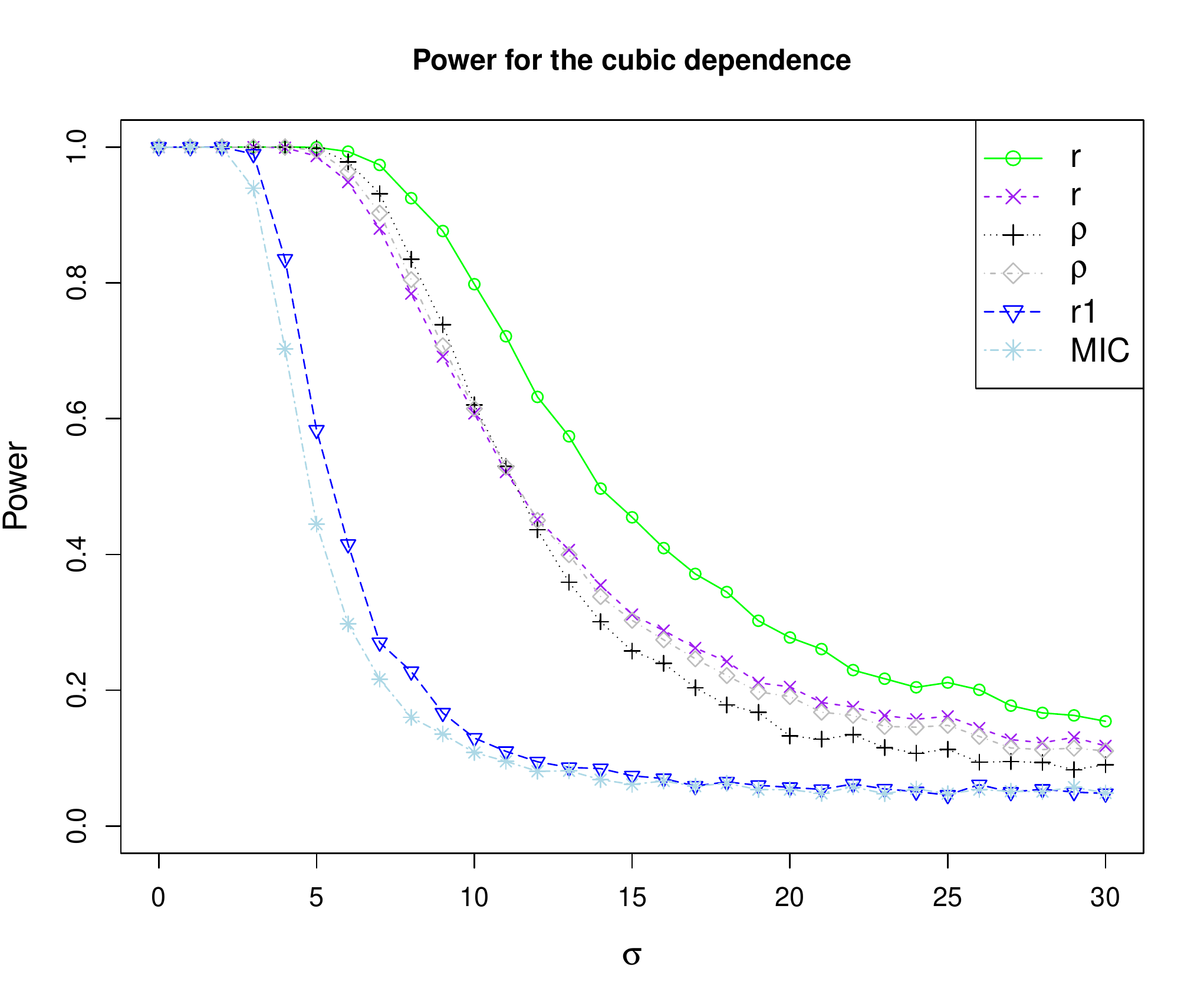}};
    \node[scale=0.9] at (4.9,4.1) {$|$};
    \node[scale=0.9] at (5.2,4.1) {$|$};
    \node[scale=0.9] at (4.9,3.6) {$|$};
    \node[scale=0.9] at (5.3,3.6) {$|$};
    \node[scale=0.7] at (5.15,3.55) {$\text{s}$};
    \node[scale=0.7] at (5.4,3.1) {$\text{max}$};
    \node[scale=0.7] at (5.4,2.8) {$\text{dist}$};
    \end{tikzpicture}
    \caption{The estimated powers of the coefficients $|r|$, $|r_s|$, $\rho_{\text{max}}$, $\rho_{\text{dist}}$, $r_1$ and MIC for $n=1000$ observations of the model \eqref{q_funcmodel} with the cubic function $f_3(x)=0.3x^3$, when $\sigma=0,1,...,30$.}
    \label{fig411}
\end{figure}
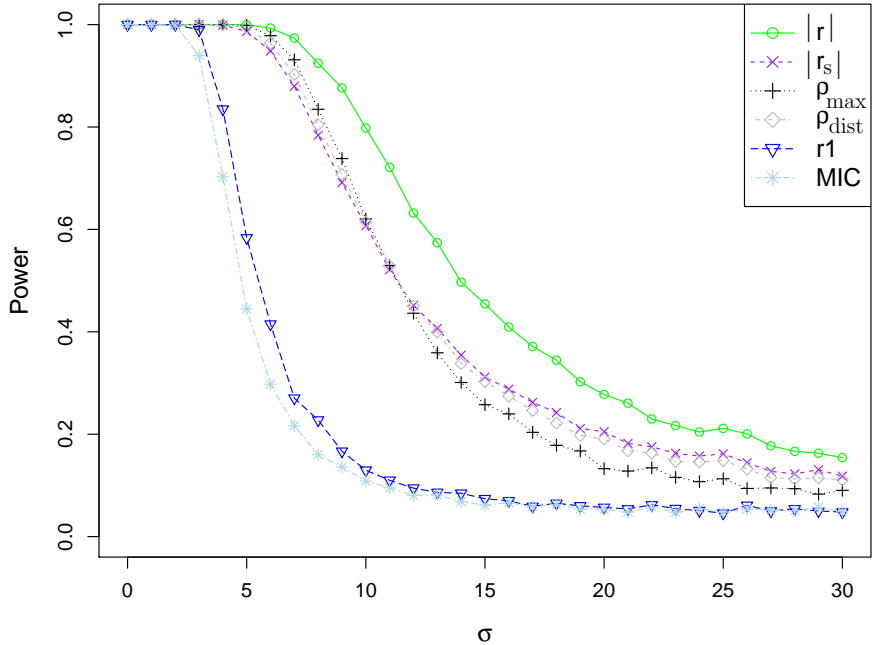

Next, let us inspect how the amount of noise affects the power of our coefficients. To do this, we first choose some model and an appropriate interval of the noise parameter $\sigma$ for this model. For each value of $\sigma$, we compute the values of our coefficients in 3000 simulations with $n=1000$ observations and estimate the powers from these results by using the critical values of Table \ref{t2} for $\alpha=5\%$. We plot the final results for three specific models.

Figure \ref{fig411} contains the powers of all our coefficients, when the model is \eqref{q_funcmodel} with the cubic function $f_3(x)=0.3x^3$ and $\sigma=0,1,...,30$. For the first few values of $\sigma$, all our coefficients have power of 1, but the powers of the MIC and the coefficient $r_1$ decrease very fast when $\sigma>3$. The most powerful measure of dependence for this model is Pearson's correlation coefficient $|r|$, followed by the coefficients $|r_s|$, $\rho_{\text{max}}$, and $\rho_{\text{dist}}$, all of whose powers seem to have very similar values.

\begin{figure}[!ht]
    \centering
    \begin{tikzpicture}
    \draw (0, 0) node[inner sep=0]
    {\includegraphics[trim={0cm 1cm 0cm 2cm},clip,width=11.5cm]{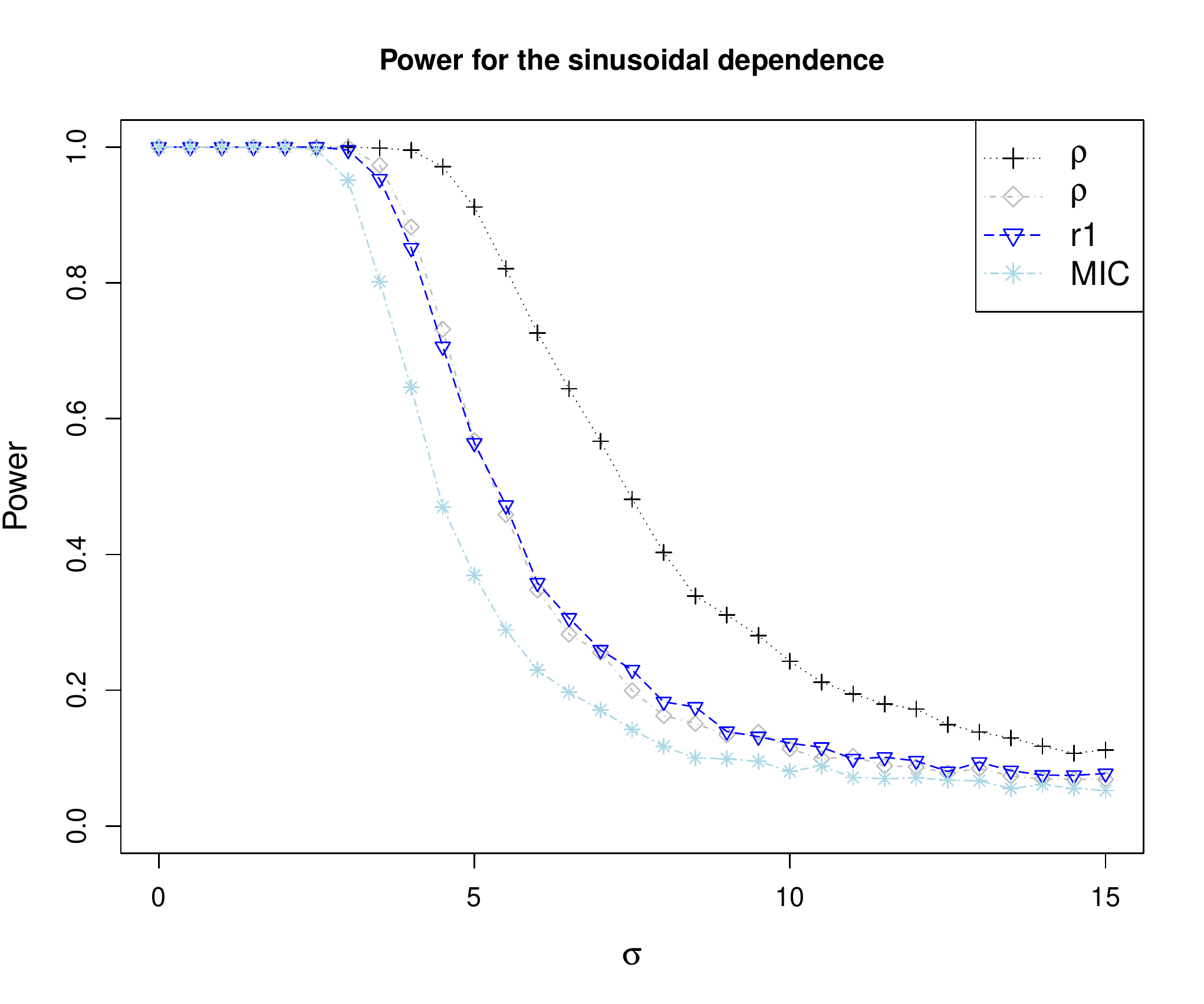}};
    \node[scale=0.7] at (4.85,3.45) {$\text{max}$};
    \node[scale=0.7] at (4.85,3.15) {$\text{dist}$};
    \end{tikzpicture}
    \caption{The estimated powers of the coefficients $|r|$, $|r_s|$, $\rho_{\text{max}}$, $\rho_{\text{dist}}$, $r_1$, and MIC for $n=1000$ observations of the model \eqref{q_funcmodel} with the sinusoidal function $f_5(x)=1.3\sin(3x)$, when $\sigma=0,0.5,...,15$.}
    \label{fig421}
\end{figure}

Let us then consider the model \eqref{q_funcmodel} but choose the sinusoidal function $f_5(x)=1.3\sin(3x)$ instead and let $\sigma=0,0.5,...,15$. Since neither Pearson's nor Spearman's correlation coefficient is well-suited for non-monotonic dependence, we only consider the coefficients $\rho_{\text{max}}$, $\rho_{\text{dist}}$, $r_1$, and MIC. From Figure \ref{fig421}, we see that the maximal correlation $\rho_{\text{max}}$ is considerably more powerful than the coefficients $\rho_{\text{max}}$ and $r_1$, whereas the MIC has the least power.

\begin{figure}[!ht]
    \centering
    \begin{tikzpicture}
    \draw (0, 0) node[inner sep=0]
    {\includegraphics[trim={0cm 1cm 0cm 2cm},clip,width=11.5cm]{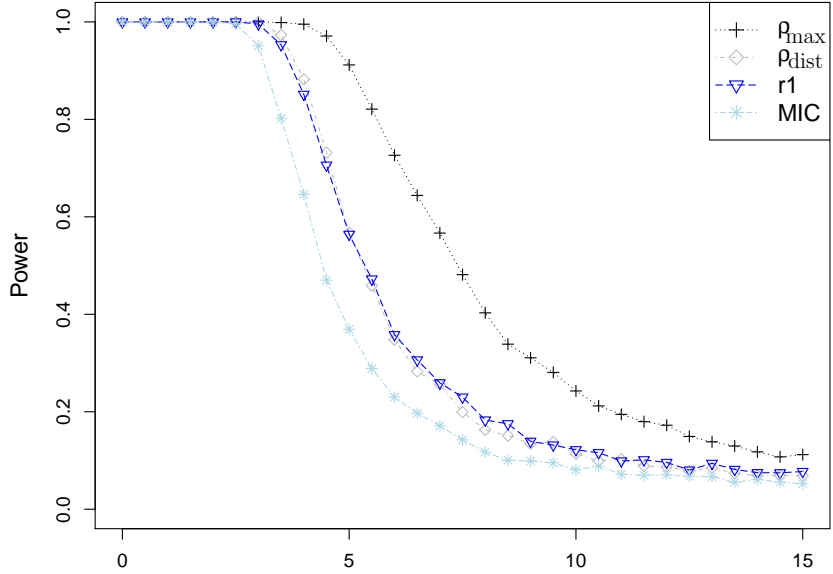}};
    \node[scale=0.7] at (4.85,3.45) {$\text{max}$};
    \node[scale=0.7] at (4.85,3.15) {$\text{dist}$};
    \end{tikzpicture}
    \caption{The estimated powers of the coefficients $|r|$, $|r_s|$, $\rho_{\text{max}}$, $\rho_{\text{dist}}$, $r_1$, and MIC for $n=1000$ observations of the cross-shaped model \eqref{q_crossmdl}, when $\sigma=0,0.1,...,3$.}
    \label{fig431}
\end{figure}

Our third model considered is cross-shaped dependence of \eqref{q_crossmdl}. Recall that $\sigma=0$ gives us here a noiseless dependence whereas $\sigma=3$ means that the data comes from two fully independent normal variables, so the powers of our coefficients should decrease from 1 to the value of $\alpha$ as $\sigma$ increases from 0 to 3. Figure \ref{fig431} is plotted by using the values $\sigma=0,0.1,...,3$ and, as we can see, the power of the MIC decreases quickly close to 0 around $\sigma=0.6$ and only the maximal correlation has values over 0.9 when $\sigma$ exceeds 1.5.

By running similar experiments for all the other models introduced in Section \ref{s3}, it can be noticed that the results found above do not change much. Namely, Pearson's correlation coefficient $|r|$ is the most powerful measure for monotonic dependence and the maximal correlation has the most power for detecting non-monotonic relationships, regardless of if they are functional or not. The MIC is very sensitive to the noise and therefore has less power than the coefficients $\rho_{\text{max}}$, $\rho_{\text{dist}}$, and $r_1$, whenever there is at least little noise in the model. This result was not affected by changing the level of significance into $10\%$ or $1\%$ with the corresponding critical values from Table \ref{t4}.

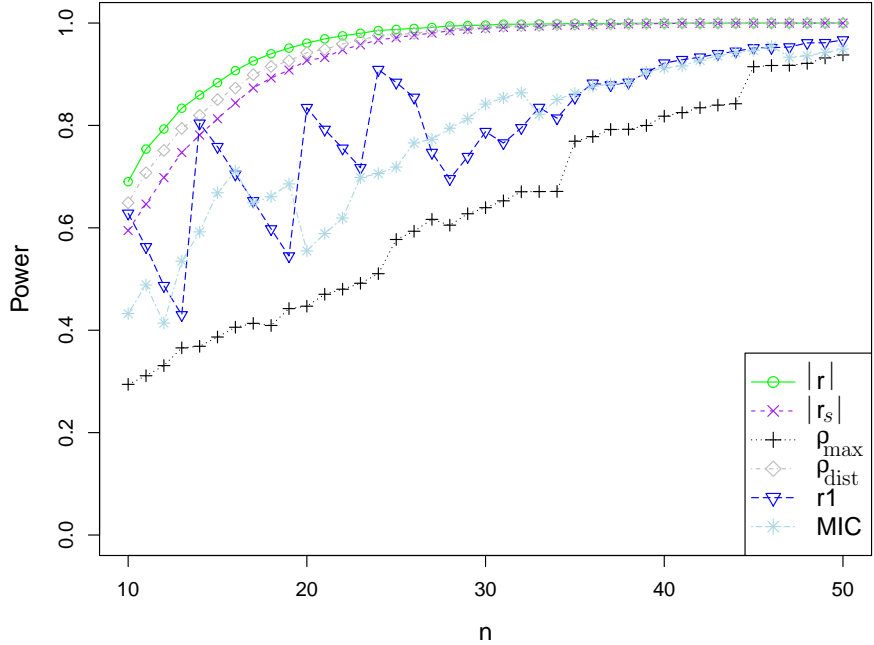
\begin{figure}[!ht]
    \centering
    \begin{tikzpicture}
    \draw (0, 0) node[inner sep=0]
    {\includegraphics[trim={0cm 0cm 1cm 2cm},clip,width=11.5cm]{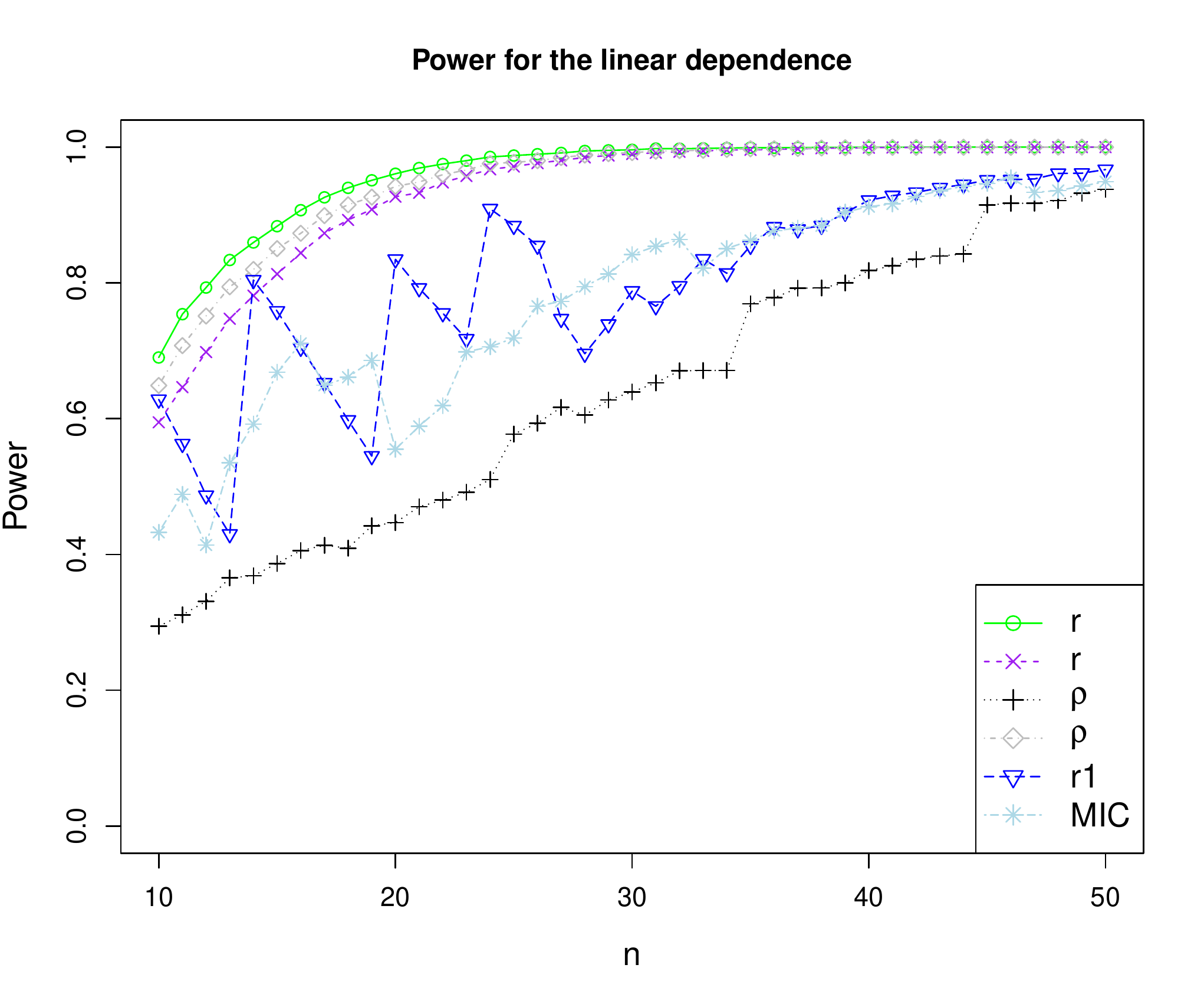}};
    \node[scale=0.9] at (4.9,-0.55) {$|$};
    \node[scale=0.9] at (5.2,-0.55) {$|$};
    \node[scale=0.9] at (4.9,-1) {$|$};
    \node[scale=0.9] at (5.3,-1) {$|$};
    \node[scale=0.7] at (5.15,-1.1) {$s$};
    \node[scale=0.7] at (5.35,-1.55) {$\text{max}$};
    \node[scale=0.7] at (5.33,-1.9) {$\text{dist}$};
    \end{tikzpicture}
    \caption{The estimated powers of the coefficients $|r|$, $|r_s|$, $\rho_{\text{max}}$, $\rho_{\text{dist}}$, $r_1$, and MIC for $n$ observations from the model \eqref{q_funcmodel} with the linear function $f_1(x)=x$ and $\sigma=1$, when $n=10,11,...,50$.}
    \label{fig441}
\end{figure}

However, if we choose the number $n$ of observations so that it is clearly less than 100, it influences on the power of the coefficients. For each $n=10,11,...,50$, we run 30000 simulations consisting of $n$ observations of two independent normal variables, use this data to compute the critical values of the coefficients with the significance level $\alpha=5\%$ and then estimate the power of these coefficients from 30000 simulations with $n$ observations from the model \eqref{q_funcmodel} where $f$ is the linear function $f_1(x)=x$ and $\sigma=1$. As can be seen from Figure \ref{fig441}, the Pearson's coefficient $|r|$ has the greatest power, followed closely by the coefficients $\rho_{\text{dist}}$ and $|r_s|$, while the maximal correlation has the least power.

Figure \ref{fig441} also shows us that the powers of the coefficient $r_1$ and the MIC are not always increasing with respect to the number $n$ of observations. This is because of our methods of computation: The mutual information needed to obtain the value of $r_1$ is estimated by using $\sqrt[3]{n}$ bins and the MIC is computed on a grid whose size is limited with the function $B(n)=\max\{n^{0.6},4\}$. By changing these default settings, we could fix this issue.

\section{Equitability for functional types of dependence}\label{s5}

In this section, we study the equitability properties of the maximal correlation, the distance correlation, the coefficient $r_1$ and the MIC. By \emph{equitability}, we mean here such feature of a measure of dependence that it gives similar values for equally noisy relationships, regardless of the exact type of the association. We focus here on the model \eqref{q_funcmodel}, where the function $f_j$ is one of the six options defined in Section \ref{s3}: linear, logarithmic, cubic, quadratic, sinusoidal, or piecewise. 

Recall the noiseless simulations of Table \ref{t1}. It is clear that neither Pearson's nor Spearman's coefficient is equitable because they do not recognize non-monotonic types of dependence so we do not consider these coefficients. Similarly, the distance correlation cannot have this property because its values vary from 0.36 to 1 for functional relationships with $\sigma=0$. Still, we can use the coefficient $\rho_{\text{dist}}$ as a control when assessing the equitability of $\rho_{\text{max}}$, $r_1$, and MIC, who all have values close to 1 for these noiseless relationships.

However, in order to inspect the impact of the noise levels on our coefficients between several models, we need such a way to measure the amount of noise that does not depend on the choice of the function $f_j$ in the model \eqref{q_funcmodel} like the previously used parameter $\sigma$ does. Consequently, we consider the \emph{coefficient of determination}, defined as \cite[p.\ 3355]{k14}
\begin{align}\label{q_cod}
R^2=R^2(f(X);Y)=(\rho(f(X);Y))^2\in[0,1],    
\end{align}
where $X$ and $Y$ are chosen so that the function $f$ defines their relationship so that $Y=f(X)+\epsilon$ with some third variable $\epsilon$ and $\rho(;)$ is the population correlation estimated with Pearson's coefficient $r$. Since the amount of noise is decreasing with respect to $R^2$, we consider here the difference $1-R^2$ instead. Note also that, according to \cite[p.\ 3355]{k14}, no non-trivial measure of dependence can be fully $R^2$-equitable, but it is still useful to know if some of our coefficients are closer to fulfilling this property than the others. 

\begin{figure}[!tbp]
  \centering
  \subfloat{\begin{tikzpicture}
    \draw (0, 0) node[inner sep=0]
    {\includegraphics[trim={0cm 0cm 0cm 0cm},clip,width=7.9cm]{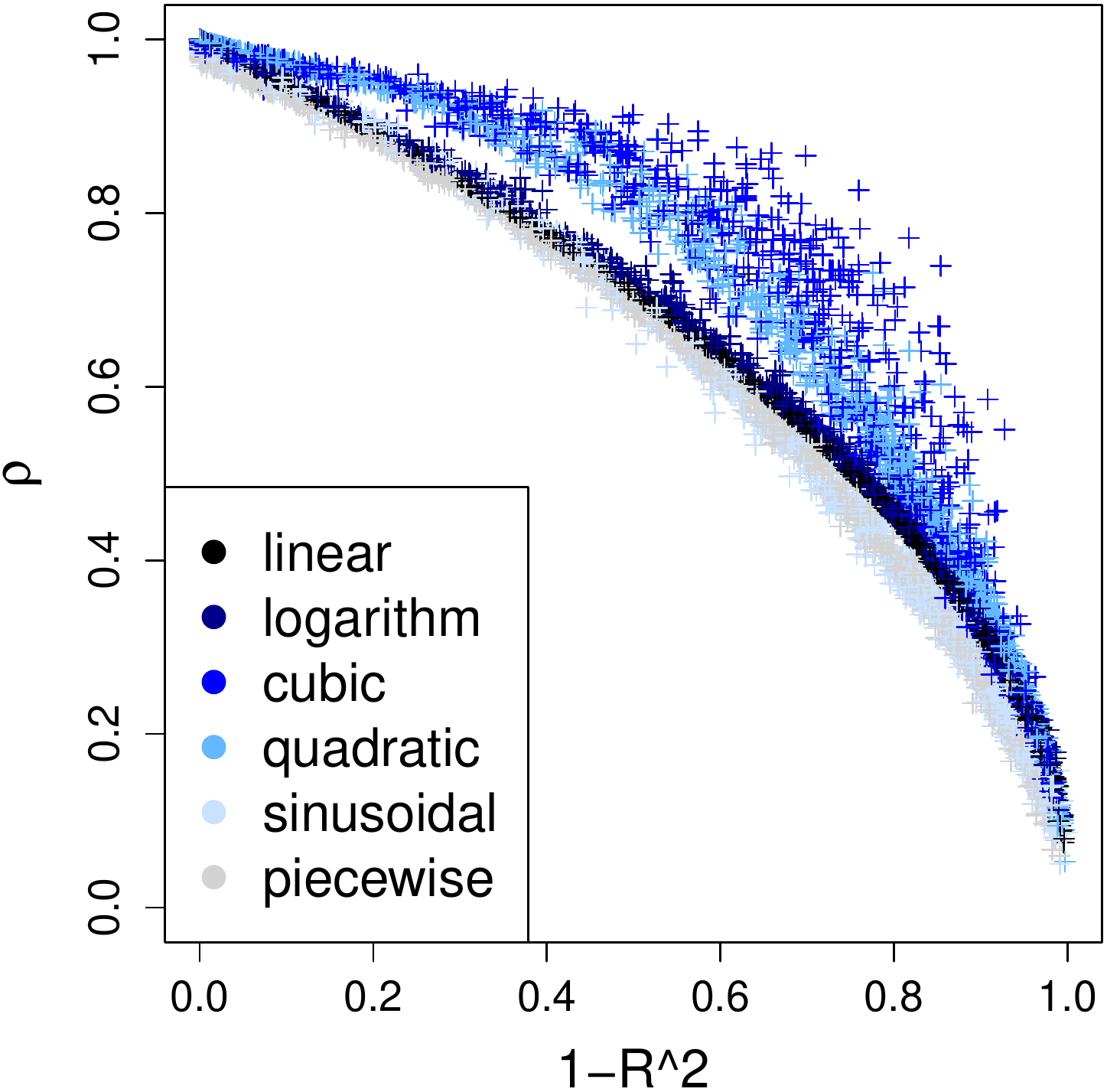}};
    \node[rotate around={90:(0,0)},scale=0.7] at (-3.6,0.8) {$\text{max}$};
    \end{tikzpicture}\label{fig511}}
  \hfill
  \subfloat{\begin{tikzpicture}
    \draw (0, 0) node[inner sep=0]
    {\includegraphics[trim={0cm 0cm 0cm 0cm},clip,width=7.9cm]{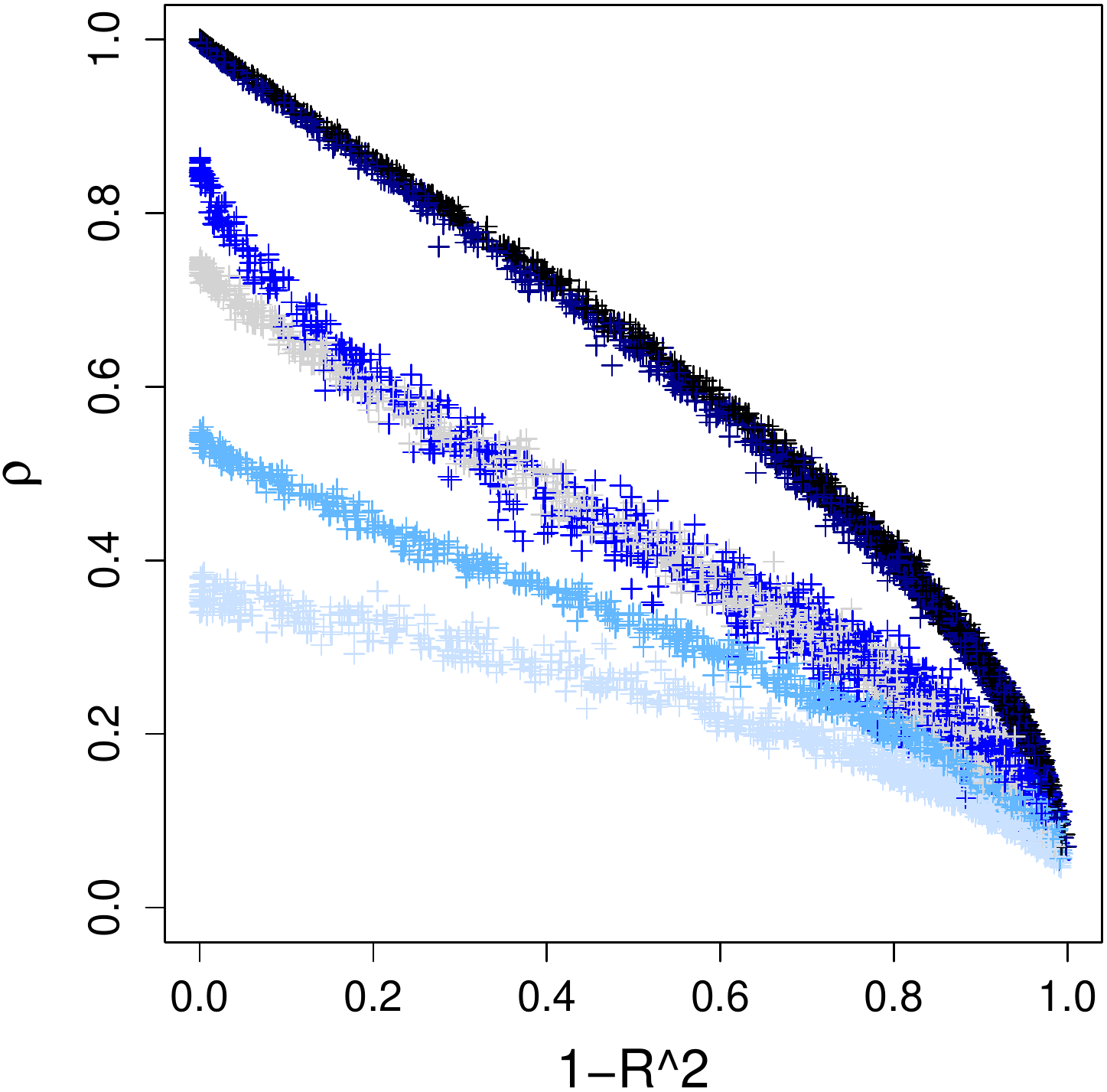}};
    \node[rotate around={90:(0,0)},scale=0.7] at (-3.7,0.8) {$\text{dist}$};
    \end{tikzpicture}\label{fig512}}
  \\
  \subfloat{\begin{tikzpicture}
    \draw (0, 0) node[inner sep=0]
    {\includegraphics[trim={0cm 0cm 0cm 0cm},clip,width=7.9cm]{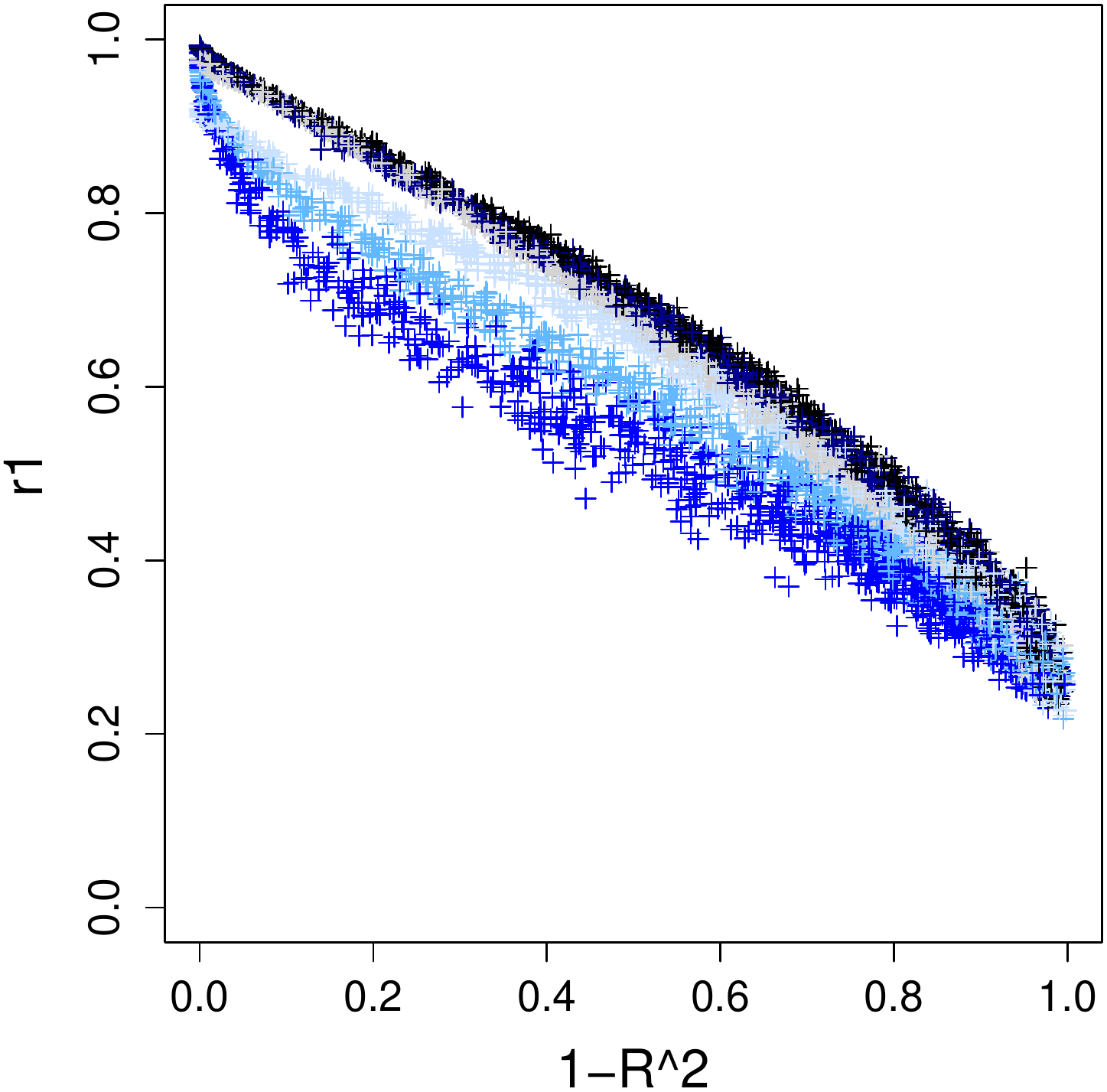}};
    \end{tikzpicture}\label{fig513}}
  \hfill
  \subfloat{\begin{tikzpicture}
    \draw (0, 0) node[inner sep=0]
    {\includegraphics[trim={0cm 0cm 0cm 0cm},clip,width=7.9cm]{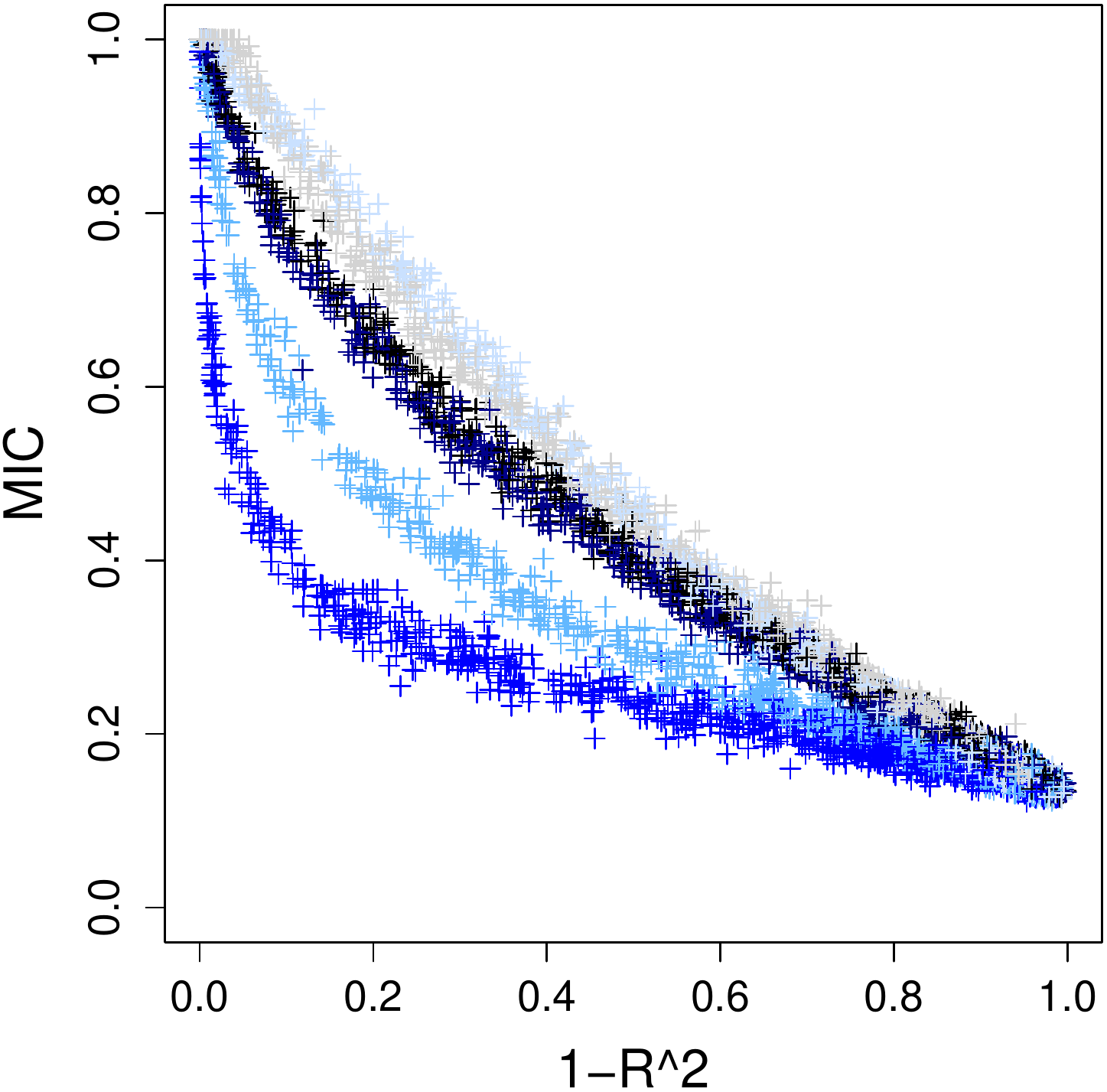}};
    \end{tikzpicture}\label{fig514}}
  \caption{The values of the coefficients $\rho_{\text{max}}$, $\rho_{\text{dist}}$, $r_1$, and MIC against the noise measured with $1-R^2$ in 1000 simulations of $n=1000$ observations from the model \eqref{q_funcmodel} with the linear, logarithmic, cubic, quadratic, sinusoidal, and piecewise functions $f_j$.}
  \label{fig5}
\end{figure}

Figure \ref{fig5} shows us how the values of each coefficient $\rho_{\text{max}}$, $\rho_{\text{dist}}$, $r_1$, and MIC change for different functional types of dependence, when the noise measured with $1-R^2$ grows. This figure was produced by generating 1000 times $n=1000$ values for $X$ and $Y$ according to the model \eqref{q_funcmodel} and, during each iteration round, computing the values of different coefficients and $1-R^2$, where $R$ is the Pearson's correlation between $f(X)$ and $Y$ obtained with the function \texttt{cor} in the R code. The results suggest that the most equitable coefficient is $r_1$, which is compatible with prior research \cite{k14} where mutual information was noted to be able to measure different types of dependence in a consistent way. The MIC fulfills here the equitable better than the distance correlation but not as well as the maximal correlation.

We also notice here one interesting aspect of the maximal correlation. Namely, for several different functions $f$ in the model \eqref{q_funcmodel}, it follows from the similarities in the definitions \eqref{q_mcc} and \eqref{q_cod} that $\rho_{\text{max}}\geq\sqrt{R^2}$. For instance, suppose that $f(X)=X$ so that our variables are $X\sim N(0,1)$ and $Y=X+\epsilon$ with $\epsilon\sim N(0,\sigma^2)$, $X\perp\epsilon$. Now, $E(X)=E(Y)=0$, $\text{Var}(X)=1$ and $\text{Var}(Y^2)=1+\sigma^2$, so by the definition of correlation,
\begin{align*}
\sqrt{R^2}&=\rho(X;Y)=\frac{E((X-E(X))(Y-E(Y))}{\sqrt{\text{Var}(X)\text{Var}(Y)}}=\frac{E(XY)}{\sqrt{1+\sigma^2}}=\frac{1}{\sqrt{1+\sigma^2}}\\
&=E\left(X\frac{Y}{\sqrt{1+\sigma^2}}\right)=\rho\left(X;\frac{Y}{\sqrt{1+\sigma^2}}\right)\leq\rho_{\text{max}},
\end{align*}
as can be visually verified from Figure \ref{fig5} even though our computational methods are not fully accurate.

The equitability cannot be directly studied for non-functional relationships because the coefficient $R^2$ is only defined for measuring noise from data that follows some functional model. Still, we know from Tables \ref{t1} and \ref{t2} that the values of the MIC are around 0.6 for both the cross-shaped dependence with no noise and the linear dependence with $\sigma\approx0.6$ or, equivalently, $R^2\approx0.7$. Since the maximal correlation has values close to 1 for all non-functional types of dependence and, unlike the MIC, this coefficient is not very sensitive to the noise, it probably has reasonably good equitability properties when measuring non-functional relationships.

\section{Conclusions}

According to our three criteria of generality, power, and equitability, the best choice of a measure of dependence is often the maximal correlation. The information coefficient of correlation $r_1$ and the distance correlation also work relatively well. However, Pearson's and Spearman's correlation coefficients are greatly limited by the type of the dependence and the MIC is not well-suited for noisy data.

Both Pearson's and Spearman's correlation coefficients can be used to recognize non-monotonic dependence also when it is non-linear, but they do not find non-monotonic dependence if it is symmetric. Surprisingly, the maximal correlation also identifies non-functional relationships, even better than the coefficients that were actually designed for this objective. The distance correlation and the coefficient $r_1$ work in an expected way but the MIC is considerably more sensitive to the amount of noise than any of the other coefficients. The number of observations does not affect very much the values of these quantities but there needs to at least 8 or so observations so that our methods of computation work properly.

For monotonic types of dependence, Pearson's correlation coefficient is the most powerful measure of dependence, regardless of the number of observations. In case of non-monotonic or non-functional dependence, the maximal correlation has the most power, assuming we have at least 100 observations in the data. If we have less than 50 observations from a non-monotonic model, the distance correlation is a good choice for a measure of dependence because it is the most powerful out of the coefficients able to recognize this association and it is not susceptible to the exact number of observations. Predictably, the power of the MIC is very weak in all cases with at least some noise when compared to the other quantities.

The coefficient $r_1$ can be used to measure functional dependence in quite an equitable way. The maximal correlation fulfills this property relatively well and, while the MIC is less equitable than the coefficient $r_1$ and the maximal correlation, it still gives values close to 1 for all functional relationships with no noise and then decreases as the amount of noise grows. In turn, the distance correlation is not equitable in any way because its values vary very much depending on the function behind the dependence, even when there is no noise.\\


\textbf{Author information.} email: \texttt{ormrai@utu.fi}, 
ORCID: 0000-0002-7775-7656,
affiliation: University of Turku, FI-20014 Turku, Finland

\textbf{Funding.} I received funding from the University of Turku Graduate School.

\textbf{Code availability.} The R code for this work is at https://github.com/oonar/til

\textbf{Conflict of interest statement.} There is no conflict of interest.

\textbf{Acknowledgements.} I am thankful to Professors Janne Kujala and Riku Kl\'en for their support during the writing process of this article, and also to the referees for their constructive and detailed comments.

\def\cprime{$'$} \def\cprime{$'$} \def\cprime{$'$}
\providecommand{\bysame}{\leavevmode\hbox to3em{\hrulefill}\thinspace}
\providecommand{\MR}{\relax\ifhmode\unskip\space\fi MR }
\providecommand{\MRhref}[2]{%
  \href{http://www.ams.org/mathscinet-getitem?mr=#1}{#2}
}
\providecommand{\href}[2]{#2}

\end{document}